\documentclass[lettersize,journal]{IEEEtran}

\usepackage{algorithmic}
\usepackage{algorithm}
\usepackage[caption=false,font=normalsize,labelfont=sf,textfont=sf]{subfig}
\usepackage{textcomp}
\usepackage{stfloats}
\usepackage{url}
\usepackage{verbatim}
\usepackage{graphicx}
\usepackage{xcolor}
\usepackage{pict2e}
\usepackage{hyperref}
\usepackage{lipsum}
\usepackage{textcomp}
\usepackage{booktabs}

\usepackage{amsmath,amsfonts}
\usepackage{dsfont}
\usepackage{mathtools}
\usepackage{bm}
\usepackage{array}
\usepackage{physics}

\usepackage{quantikz}

\usepackage[sorting=none, maxbibnames=2, giveninits=true, date=year, doi=false,isbn=false,url=false,eprint=false, backend=biber]{biblatex}
\AtEveryBibitem{\clearfield{pages}} 

\DeclareMathOperator{\opspan}{span}

\DeclareMathOperator{\optrace}{Tr}
\newcommand{\RR}[0]{\mathbb{R}}
\newcommand{\CC}[0]{\mathbb{C}}
\newcommand{\calH}[0]{\mathcal{H}}
\newcommand{\calX}[0]{\mathcal{X}}
\newcommand{\calY}[0]{\mathcal{Y}}

\newcommand{\red}[1]{\textcolor{black}{#1}}

\newsavebox{\ORCIDlogo}
\savebox{\ORCIDlogo}{%
\setlength{\unitlength}{\dimexpr 1em/256\relax}%
\begin{picture}(256,256)%
  \color[HTML]{A6CE39}\put(128,128){\circle*{256}}%
  \color{white}%
  \put(78.6,199.2){\circle*{20}}%
  \moveto(70.9,176,9)\lineto(86.3,176,9)\lineto(86.3,69.8)\lineto(70.9,69.8)%
  \closepath\fillpath%
  \moveto(108.9,176.9)\lineto(150.5,176.9)%
  \curveto(190.1,176.9)(207.5,148.6)(207.5 ,123.3)%
  \curveto(207.5,95,8)(186,69.7)(150.7,69.7)%
  \lineto(108.9,69.7)%
  \closepath\fillpath%
  \color[HTML]{A6CE39}%
  \moveto(124.3,83.6)\lineto(148.8,83.6)%
  \curveto(183.7,83.6)(191.7,110.1)(191.7,123.3)%
  \curveto(191.7,144.8)(178,163)(148,163)%
  \lineto(124.3,163)%
  \closepath\fillpath%
\end{picture}%
}

\newcommand\orcidicon[1]{\href{https://orcid.org/#1}{\usebox{\ORCIDlogo}}}

\begin{document}

\title{Automatic and effective discovery of quantum kernels}

\author{
    \IEEEauthorblockN{
    Massimiliano~Incudini\IEEEauthorrefmark{1}\IEEEauthorrefmark{4}\orcidicon{0000-0002-9389-5370}, 
    Daniele~Lizzio~Bosco\IEEEauthorrefmark{2}\orcidicon{0009-0002-7372-6518}, 
    Francesco~Martini\IEEEauthorrefmark{1}\orcidicon{0000-0003-2651-140X}, \\
    Michele~Grossi\IEEEauthorrefmark{3}\orcidicon{0000-0003-1718-1314}, 
    Giuseppe~Serra\IEEEauthorrefmark{2}\orcidicon{0000-0002-4269-4501}, 
    Alessandra~Di~Pierro\IEEEauthorrefmark{1}\orcidicon{0000-0003-4173-7941}
    }
    
    \IEEEauthorblockA{\IEEEauthorrefmark{1}Department of Computer Science, University of Verona, Verona 37134, Italy}
    
    \IEEEauthorblockA{\IEEEauthorrefmark{2}Department of Computer Science, University of Udine, Udine 33100, Italy}
    
    \IEEEauthorblockA{\IEEEauthorrefmark{3}European Organization for Nuclear Research (CERN), Geneva 1211, Switzerland}
    
    \IEEEauthorblockA{\IEEEauthorrefmark{4}Correspondence to: {massimiliano.incudini@univr.it}}
}

\markboth{}{}

\maketitle

\begin{abstract}
Quantum computing can empower machine learning models by enabling kernel machines to leverage quantum kernels for representing similarity measures between data. Quantum kernels are able to capture relationships in the data that are not efficiently computable on classical devices. However, 
there is no straightforward method to engineer the optimal quantum kernel for each specific use case.
We present an approach to this problem, which employs optimization techniques, similar to those used in neural architecture search and AutoML, to automatically find an optimal kernel in a heuristic manner. To this purpose we define an algorithm for constructing a quantum circuit implementing the similarity measure as a combinatorial object, which is evaluated based on a cost function and then iteratively modified using a meta-heuristic optimization technique. The cost function can encode many criteria ensuring
favorable statistical properties of the candidate solution, such as the rank of the Dynamical Lie Algebra. Importantly, our approach is independent of the optimization technique employed.
The results obtained by testing our approach on a high-energy physics problem demonstrate that, in the best-case scenario, we can either match or improve testing accuracy with respect to the manual design approach, showing the potential of our technique to deliver superior results with reduced effort. 
\end{abstract}


\section{Introduction}

In recent years, there has been significant progress in machine learning, leading to advancements in fields like Computer Vision, Natural Language Processing, and Predictive Analytics. Tools like Dall-E3 and ChatGPT have the potential to revolutionize society. Nonetheless, certain applications remain beyond the reach of classical means, and these might benefit from the use of quantum machine learning (QML) 
\cite{biamonte2017quantum, cerezo2022challenges}. 
One technique that has garnered significant interest is \emph{quantum kernel} \cite{schuld2019qmlfeaturehilbert}, not least due to its similarities with classical kernel methods—a popular approach in statistical learning \cite{steinwart2008svmbook}. The latter have already demonstrated their effectiveness by enhancing supervised and unsupervised models in various fields, such as image classification, audio processing, and text analysis \cite{muandet2017kernel}.

Kernel methods take into account the geometric and topological aspects of the data, enabling the identification of complex patterns and connections within the dataset. In particular, they allow the generalization of the conventional Euclidean product by using as a kernel any arbitrary positive semi-definite function.
Quantum kernels share the mathematical framework of classical kernel methods but, contrary to the classical kernels, are implemented as linear operators on the Hilbert space of a quantum system whose states encode the classical (or quantum) data. The feasibility of this approach was originally proved experimentally in \textcite{havlivcek2019supervised}.

Quantum kernels with a computational 
speedup over classical methods can be defined by assuming the availability of fault-tolerant quantum computers on which  algebraic subroutines solving NP problems can be run using the full power of the quantum computation paradigm. For example, this is the approach taken in \cite{incudini2023higher} for defining a quantum kernel based on the (classically infeasible) computation of higher order Betti numbers, and for achieving a proven advantage on some hard-to-compute artificial learning tasks based on the prime factorization problem \cite{liu2021rigorous} and BQP-complete problems \cite{jager2023universal}.

In contrast, the definition of quantum kernels that are more amenable to be run on NISQ devices, i.e. the current generation of quantum hardware \cite{preskill2018quantum},
has found applications in physics. For example, in the field of Particle Physics, \textcite{wozniak2023quantum} proposes an anomaly detection method based on a quantum kernel,
which determines whether a given set of features extracted from collision events belong to the Standard Model or not, showing a consistent performance improvement over classical methods.

Although these latter quantum kernels may reach an advantage by means of relatively shallow quantum circuits compared to those required by the fault-tolerant approach, finding a quantum kernel suitable for a specific task is not straightforward. Firstly, the expressivity of the quantum kernel, i.e. the ability of a kernel to map data points into any region of the Hilbert space, influences its performance. Highly expressive kernels suffer from an exponential concentration of kernel values, requiring an exponential amount of samples to distinguish them \cite{thanasilp2022exponential}. Moreover, a quantum kernel with high expressivity may have a low learning power,
since when decomposing the kernel into orthonormal components, the corresponding eigenvalues are exponentially small in the number of qubits $n$. Since each component can be learned with a sample size proportional to the corresponding eigenvalue, this results in a quantum kernel incapable of learning any target function \cite{kubler2021inductive}. This problem can be mitigated by using projected quantum kernels \cite{huang2021projected, gan2023unified}, which project the data to a smaller Hilbert space, or by limiting the mapping of data to a small region of the exponentially large Hilbert space \cite{canatar2022bandwidth}. Another solution is shown in \cite{ragone_representation_2023}, where techniques from the domain of geometric deep learning \cite{bronstein_geometric_2021} are used when the dataset exhibits specific symmetries, as is the case for image-related tasks that exhibit invariance to reflection and rotation. Although this approach yields interesting results, exploiting the symmetries of a dataset requires specific knowledge of the application domain, which is often hard to identify. Secondly, it is difficult to design a quantum kernel that captures the inductive bias and the set of assumptions on the nature of the target function, which are necessary to achieve well-performing models. Therefore, it is important to undertake the challenging task of designing effective quantum kernels in a general, unknown setting. 

In this paper, we propose an approach to automatically construct quantum kernels that are tailored to the specific features of a given task.  
This construction is based on combinatorial optimization techniques and encompasses two key aspects. First, a quantum kernel is represented as a combinatorial object whose modeling places special emphasis on limiting the expressivity of the unitary matrix.
Second, a heuristic algorithm is employed to iteratively explore the space of quantum kernels, taking into account the performance of the kernel on the given application scenario or other properties the candidate solution should exhibit (e.g. the degree of expressivity).
In proposing this approach, we were inspired by the success of \emph{neural architecture search}, \emph{AutoML}, and \emph{symbolic regression} techniques in classical machine learning. A comparison of our approach with the existing literature is presented in the Supplementary Material~S-I. 

We demonstrate our method for automatic kernel discovery on a problem in High Energy Physics (HEP), namely anomaly detection in proton collision events. 
This task is pivotal for the advancement of our understanding of new phenomena, which could help unraveling some of the persistent questions in the Standard Model of particle physics, a central theme in the LHC (Large Hadron Collider) physics programme objectives.
To evaluate the performance of the discovered kernels, we compare our results with classical techniques as well as with the kernels presented in 
\cite{wozniak2023quantum}, which is currently the best-performing quantum machine learning algorithm for the task under the specified data compression condition. To ensure a fair comparison, we maintained the same experimental setup and machine learning model used in their study, with the sole modification being the quantum kernel.

Our results showcase the capability of our approach to achieve performances that match or surpass the results obtained via a manual design of quantum kernels, by using an automated selection tool based on optimization.

The paper is structured as follows. Section~\ref{sec:preliminaries} introduces the necessary background, while Section~\ref{sec:methods} introduces our approach showing how to model quantum kernels as combinatorial objects, how their fitness for specific tasks can be evaluated and their favorable statistical properties recovered, and finally how to optimize those combinatorial objects. In Section~\ref{sec:application}, we apply our approach to the proton collision anomaly detection problem and compare our results with the state-of-the-art literature. Finally, in Section~\ref{sec:conclusion} we discuss the promises and limitations of our approach and suggest future directions.

\section{Preliminaries}\label{sec:preliminaries}

We briefly introduce the necessary background and refer for a more detailed treatment to \cite{steinwart2008svmbook} for a classical background, \cite{manentimotta2023quantumbook} for a quantum background, and \cite{schuld2021qmlbook,incudini2024toward} for quantum kernels.

\subsection{Kernel methods}

\red{Let $\{(\bm{x}^{(j)}, y^{(j)}) \}_{j=1}^p \subseteq \calX \times \calY$ be a labelled dataset of pairs that are i.i.d. sampled from an unknown probability distribution. We indicate the data vector space with $\calX = \RR^d$, and the target space with either $\calY = \RR$ or $\calY \subseteq \mathbb{Z}, |\calY| < \infty$ for regression or classification tasks, respectively.} A \emph{kernel function} $\kappa: \calX \times \calX \to \RR$ over a non-empty set $\calX$ is defined as a function
\begin{equation}
\kappa(\bm{x}, \bm{x}') = \langle \varphi(\bm{x}), \varphi(\bm{x}') \rangle_{\mathcal{H}},
\end{equation}
where $\varphi: \calX \to \calH$ is a \emph{feature map}, i.e. a function that maps data from the original space $\calX$ to a Hilbert space $\calH$ with inner product $\langle \cdot, \cdot \rangle_\calH$. 
Kernel functions generalize the notion of similarity between data. The feature map extracts and highlights important patterns in the data, leading to an improvement in the model's performance. 
Mercer's theorem shows that any positive semidefinite $\kappa$ is a kernel function for some $\varphi$ and $\calH$, and it allows for an eigendecomposition in terms of non-negative eigenvalues and orthonormal (in $\calH$) eigenfunctions,
\begin{equation}\label{eq:mercer}
    \kappa(\bm{x}, \bm{x}') = \sum_{j=0}^\infty \lambda_j \phi_j(\bm{x}) \phi_j(\bm{x}').
\end{equation}

A Reproducing Kernel Hilbert Space (RKHS) on $\calX$ is denoted as
\begin{equation*}
    \mathcal{R} = \overline{\opspan}\Big\{ \kappa_{\bm{x}} \mid \bm{x} \in \calX, \langle \kappa_{\bm{x}}, \kappa_{\bm{x}'} \rangle = \kappa(\bm{x}, \bm{x}') \Big\}
\end{equation*}
where $\kappa$ is a reproducing kernel of $\mathcal{R}$, i.e. $\kappa_{\bm{x}} = \kappa(\bm{x}, \cdot)$. Furthermore, $\overline{\opspan}(S)$ denotes the closure of $\opspan(S)$, completing the pre-Hilbert space $\opspan(S)$ to a Hilbert space. 
The \emph{Representer theorem} \cite{steinwart2008svmbook} guarantees that the function $\tilde{f} \in \mathcal{R}$ that minimizes the (regularized) empirical risk \red{over the training set} is found via a convex, $p$-dimensional optimization problem whose solution is in the form
\begin{equation}\label{eq:kernelmachine}
\tilde{f}(\cdot) = \sum_{j=1}^p \bm{\alpha}_j \, \kappa(\bm{x}^{(j)}, \cdot).
\end{equation}

Among the different algorithms to train a model in Equation \ref{eq:kernelmachine}, the support vector machine (SVM) is among the most popular. This algorithm leads to sparse solutions, meaning that the function will have only a few non-zero weights $\bm{w}_j$. While SVMs are commonly used for classification tasks, there is a variant known as the one-class SVM (OC-SVM). The OC-SVM is trained on data from a single class and is designed to identify outliers or novel data points \cite{scholkopf1999support}. For a detailed derivation of this model, please refer to 
Supplementary Material~S-II.

\subsection{Quantum computing}

\red{The state of a} quantum system of $n$-qubits is \red{an element of the} Hilbert space 
\begin{equation}
\calH = \{ \rho \in \CC^{2^n \times 2^n} \mid \rho \ge 0, \;\optrace[\rho] = 1\};
\end{equation}
$\rho$ is a \emph{pure state} if it can be expressed as a projector, $\rho = \ketbra{\psi}{\psi}$, with $\ket{\psi} \in \CC^{2^n}$ being a unitary vector and $\bra{\psi} = (\ket{\psi})^\dagger$. In our context, $\calH$ will denote both the quantum system's space and the target space for the feature map, as they coincide. The computation is carried out by evolving the state using a unitary operator $U$, i.e. an operator satisfying $UU^\dagger = U^\dagger U = I$. If this operator depends on one or more parameters $\bm{\theta}$ then is denoted as $U(\bm{\theta})$. 

Information from a quantum system is extracted using a projection-valued measure (PVM),
\begin{equation}
    M = \{ \Pi_j \ge 0 \mid \Pi_j = \Pi_j^\dagger, \; \Pi_i \Pi_j = \delta_{i,j} \}_{j=1}^v,
\end{equation}
satisfying $\sum_{j=1}^v \Pi_j = I$. The outcomes $j$ of a measurement are observed with probability $\optrace[\rho \Pi_j]$. The measurement is destructive and, after its application, the quantum state collapses to $\Pi_j\rho \Pi_j^\dagger / \optrace[\Pi_j\rho]$. The measurement in the computational basis corresponds to the PVM $\{ \ketbra{j}{j} \}_{j=0}^{2^n-1}$. Given a quantum state $\rho$ and a Hermitian operator (or observable) $O$, measuring $O$ in the state $\rho$ corresponds to calculating the expectation value $\optrace[\rho O]$. There are more general types of measurements available, but they are not used in this work.

A unitary operation can be described in terms of a quantum circuit on $n$ wires, with the fan-in (number of inputs) matching the fan-out (number of outputs). The quantum gates implementing operators in a \emph{universal set}, whose elements are called elementary gates, usually correspond to the Pauli gates $\sigma_1 = \sigma_\textsc{x} = \smqty(0 & 1 \\ 1 & 0), \sigma_2 = \sigma_\textsc{y} = \smqty(0 & -i \\ i & 0 ), \sigma_3 = \sigma_\textsc{z} = \smqty(1 & 0 \\ 0 & -1)$, $\sigma_0 = I = \smqty(1 & 0 \\ 0 & 1)$, the CNOT gate $\textsc{cnot} = \ketbra{0}{0} \otimes I + \ketbra{1}{1} \otimes \sigma_\textsc{x}$, the rotational single- and two-qubit gates $R_\alpha(\theta) = \exp(-i\frac{\theta}{2} \sigma_\alpha)$ and $R_{\alpha\beta}(\theta) = \exp(-i\frac{\theta}{2} \sigma_\alpha \otimes \sigma_\beta)$, with $\alpha, \beta \in \{0, 1, 2, 3\}$.

\subsection{Quantum kernels methods}

We can encode classical data $\bm{x} \in \RR^d$ in the Hilbert space of a quantum system by evolving the state using a unitary operator (and underlying circuit) $U(\bm{x})$,
\begin{equation}
\varphi(\bm{x}) = U(\bm{x}) \rho_0 U(\bm{x})^\dagger,
\end{equation}
where $\rho_0$ represents the initial state of the quantum system. Then,
\begin{equation}\label{eq:quantum_kernel}
\kappa(\bm{x}, \bm{x}') = \langle \varphi(\bm{x}), \varphi(\bm{x}') \rangle_\calH = \optrace[\rho_{\bm{x}} \rho_{\bm{x}'}],
\end{equation}
is a kernel function. \red{Here, quantum states are described using the more general notion of mixed state $\rho$ even though the evolution is purely unitary, as this notation allows a more direct interpretation of quantum kernels are kernel functions, c.f. \cite{schuld2021qmlbook}.} Equation \ref{eq:quantum_kernel} can be implemented using the \emph{overlap test}, whose underlying quantum circuit (c.f. Figure \ref{fig:circuit_overlap_test}) implements the transformation:
\begin{equation}\label{eq:overlap_quantum_kernel}
\kappa(\bm{x}, \bm{x}') = \optrace[(U(\bm{x}') U^\dagger(\bm{x}) \rho_0 U(\bm{x}) U^\dagger(\bm{x}')) \ketbra{0}{0}].
\end{equation}

The \emph{expressivity} of $U(\bm{x})$ can be intuitively understood as its capacity to map classical data into quantum states within the exponentially vast Hilbert space \cite{sim2019expressibility}. To formally quantify the expressivity, we employ the norm of the super-operator $A$, 
\begin{equation}\label{eq:norm_A}
A = \int_\text{Haar} (\ketbra{\phi}{\phi})^{\otimes t} d\phi - \int_\Theta (U(\bm{\theta}) \ketbra{0}{0} U^\dagger (\bm{\theta}))^{\otimes t} d\bm{\theta},
\end{equation}
with $t \ge 2$. This formula quantifies how the ensemble of states generated by randomly sampling $\bm{\theta}$ deviates from a state $t$-design, an ensemble of states that closely approximates the Haar random ensemble of states up to the $t$-th statistical moment \cite{nakata2014generating}.

\section{Automatic discovery of quantum kernels}\label{sec:methods}

\begin{figure}[tbp]
    \centering
    \includegraphics[width=0.85 \linewidth]{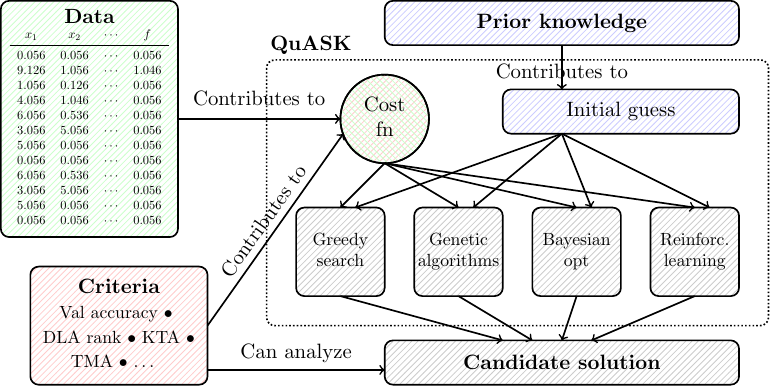}
    \caption{Pipeline for the automatic discovery of quantum kernels: Prior knowledge of the problem can be employed to propose a sub-optimal educated guess as an initial candidate solution. The cost function that evaluates the quantum kernel at each iteration is based on criteria proposed in the literature and can depend on the problem's data. Various optimization algorithms can be employed to reach the candidate solutions. The same criteria used for the cost function can also be used to further evaluate the candidate solution and explain its performance.}
    \label{fig:platform_overview}
\end{figure}

In this section, we introduce our algorithm for the automatic construction of a quantum kernel. \red{Firstly, we model an $n$-qubit, $m$-gates quantum kernel as a discrete object, $\bm{k}$, in the space,
$\mathcal{QK}_{n,m}$, of $n$-qubits, $m$-gates quantum kernels. The elements of these sets can be represented as vectors of integers in $\mathbb{N}^{6m+n}$. A vector in $\mathbb{N}^{6m+n}$ encodes the quantum kernel over $n$ qubits, with a parameterized quantum circuit containing $m$ gates. The parameterized quantum circuit itself is described by a matrix of $6 \times m$ features, which means 6 features per operation. Measurement is applied to a subset of the $n$ target qubits, contributing to the last $n$ features.}
Secondly, we introduce a family of criteria denoted by $\mathcal{C} = \{ C_1, C_2, \ldots \}$, where each $C_i: \mathcal{QK}_{n,m} \to \RR_{\ge 0}$. These criteria individually quantify various aspects of a quantum kernel, such as expressivity, efficiency of classical simulability, or compatibility with the task at hand. Each of these criteria corresponds to specific theoretical properties documented in the literature. 
The criteria will be used to define a cost function in terms of a (possibly non-linear) map of these
\begin{equation}\label{eq:generic_cost_function}
C(\bm{k}) = f(C_1(\bm{k}), C_2(\bm{k}), \ldots).
\end{equation}
Thirdly, we select an initial guess. This initial guess can be informed by prior knowledge of the task, where we speculate that a candidate solution may be close to a satisfactory one. Alternatively, it can be the identity circuit. Finally, we employ a heuristic optimization procedure to solve the problem:
\begin{equation}\label{eq:optimization}
\bm{k}^* = \arg\min_{\bm{k} \in \mathcal{QK}_{n,m}} C(\bm{k}).
\end{equation}
The use of a heuristic procedure is forced by the nonconvexity of the cost function and, like all heuristics, gives no guarantee of finding an optimal solution. Our scheme and the underlying software we have developed allow for the use of different heuristics, and is depicted in Figure~\ref{fig:platform_overview}. 

\subsection{Modeling quantum kernels}

The method we describe generalizes the approach shown in \cite{ostaszewski2021reinforcement}. Consider the quantum circuit implementing a feature map of $d$-dimensional data, $\bm{x} \in \RR^d$, which operates on $n$ qubits and consists of $m$ gates. A matrix $\mathbb{N}^{6m}$ is used to represent an $n$-qubit, $m$-gate parameterized quantum circuit \red{only ($n$ additional features will be used to describe the measurement)}. \red{This representation is convenient as the quantum circuit itself is inherently a combinatorial object, making the discrete matrix form a natural choice for modeling parametric quantum circuits.} In the matrix, each column represents the application of a single- or two-qubit gate. Each gate takes the form:
\begin{equation}
G = \exp\left(-i \frac{b_j \bm{x}_k}{2} \sigma_\alpha^{(q)} \sigma_\beta^{(q')}\right) = R_{\alpha, \beta}^{(q, q')}(\bf{b}_j \, \bm{x}_k).
\end{equation}
Here, $\bm{x}_k$ represents the $k$-th feature of the data point $\bm{x} \in \RR^d$, and $\sigma_\alpha$ and $\sigma_\beta$ are the Pauli generators of the quantum gate, $\sigma_\alpha, \sigma_\beta \in \{ \sigma_0, \sigma_1, \sigma_2, \sigma_3\}$. When $\sigma_\alpha = \sigma_0$ or $\sigma_\beta = \sigma_0$, the gate effectively operates on a single qubit. The pair $(q, q')$, where $q \neq q'$, specifies the qubits on which the quantum gate acts. Additionally, $0 < b_j \le 1$ is an optional scaling factor called \emph{bandwidth}. The possible bandwidth values belong to a discrete set $\bm{b} = \{ b_1, ..., b_B = 1\}$. Each operation is described by a tuple:
\begin{equation}
(\alpha, \beta, q, q', k, j)
\end{equation}
Here, $\alpha, \beta \in \{0,1,2,3\}$ indicates one of the four Pauli matrices, $q \in \{0, ..., n-1\}$ for the first qubit on which the transformation acts, $q' \in \{0, ..., n-2\}$ for the second qubit, which cannot be equal to the first one, $k \in \{0, ..., d-1\}$ for the feature parameterizing the transformation, and $j \in \{0, ..., B-1\}$ for the bandwidth value. Finally, it is worth noting that some works do not use features $\bm{x}_k$ directly as the rotational angle of a parameterized gate. Instead, they prefer using a function of one feature or multiple features. For example, the work in \cite{havlivcek2019supervised} uses angles $\theta_{k, k'} = (\bm{x}_k - \pi)(\bm{x}_{k'} - \pi)$. Such a result can also be obtained in our framework by adding the engineered features as new components to the feature vector $\bm{x}$. 

With the given information, we can already implement the most basic kind of quantum kernel, which is based on the `overlap test' (cf. Equation~\ref{eq:overlap_quantum_kernel}). The underlying feature map is defined as $U(\bm{x}) = \prod_{g = 1}^m R_{\alpha_g, \beta_g}^{(q_g, q_g')}(\beta_{j_g} \bm{x}_{k_g})$, and is used to encode the two data points $\bm{x}$ and $\bm{x}'$ for which we want to calculate the kernel. The corresponding circuit is depicted in Figure \ref{fig:circuit_overlap_test}. To calculate the inner product, we measure in the computational basis and estimate the probability of obtaining the measurement outcome $\ketbra{0^n}{0^n}$.

\begin{figure}[tbp]
    \centering
\scalebox{0.7}{\begin{quantikz}
    \lstick{$\ket{0}^{\otimes n}$} 
    & \gate{U(\bm{x}^{(1)})} 
    & \gate{U^\dagger(\bm{x}^{(2)})}
    & \meter{}
\end{quantikz}}
    \caption{The `overlap test' quantum circuit used to estimate the inner product of two vectors $\bm{x}^{(1)}, \bm{x}^{(2)}$ encoded via the unitary transformation $U$. The probability of measuring $0^n$ as the output of the quantum circuit with infinite precision (shots) is estimated by the means of the formula in Equation~\ref{eq:overlap_quantum_kernel}.}
    \label{fig:circuit_overlap_test}
\end{figure}

\begin{figure}[tbp]
    \centering
    \scalebox{0.7}{\begin{quantikz}[row sep=3.5]
\lstick{$\ket{0}$} 
    & \gate{H}
    & \ctrl{2}
    & \gate{H}
    & \meter{} \\
\lstick{$\ket{0}^{\otimes n}$} 
    & \gate{U(\bm{x}^{(1)})}
    & \targX{}
    & 
    &  \\
\lstick{$\ket{0}^{\otimes n}$} 
    & \gate{U(\bm{x}^{(2)})} 
    & \targX{}
    & 
    & 
\end{quantikz}}
    \caption{The `swap test' quantum circuit used to estimate the inner product of two vectors $\bm{x}^{(1)}, \bm{x}^{(2)}$ encoded accordingly to the unitary transformation $U$. It uses $2n+1$ qubits, instead of the $n$ qubits used by the overlap test, but results in shallower circuits. The probability of measuring 0 on the topmost qubit allows us to estimate the inner product via Equation~\ref{eq:swap_test}. By applying the controlled operation of the swap test to a subset of the $n$ target qubits of each data-related state, we can calculate partial inner products which can be interpreted as a projected quantum kernel (cf. Equation~\ref{eq:overlap_quantum_kernel}).}
    \label{fig:circuit_cswap_test}
\end{figure}

An alternative procedure to estimate the inner product is the `swap test', whose circuit is depicted in Figure~\ref{fig:circuit_cswap_test}. The swap test operates by estimating the probability of obtaining a zero outcome from the topmost auxiliary qubit, with the inner product being a linear function of this outcome, 
\begin{align}\label{eq:swap_test}
    & \kappa(\bm{x}, \bm{x}') = 2\optrace[V \rho_0 V^\dagger O] - 1 \\
    & V = (H \otimes I_n \otimes I_n) (c\textsc{swap}) (H \otimes U(\bm{x}) \otimes U(\bm{x}')) \nonumber \\
    & O = \ketbra{0}{0} \otimes I_n \otimes I_n \nonumber
\end{align}
When the controlled-SWAP gate acts on a subset of the $n$ target qubits of the two states holding the data, it will result in a partial measurement. 

The possibility of performing partial inner products allows us to test similarity within a significantly smaller Hilbert space compared to the $n$-qubit space, resulting in a kernel function with reduced expressivity. This, in turn, can lead to better generalization performance, as will be discussed in Section~\ref{sec:expressibility_control}. A string $s = \{0,1\}^n$ indicates which qubits need to be measured, with $s_i = 1$ if the $i$-th qubit has to be measured. This method exhibits similarities with the \emph{projected quantum kernel} introduced in \cite{huang2021projected}. 

\begin{figure}[tbp]
    \centering
    \includegraphics[width=0.85\linewidth]{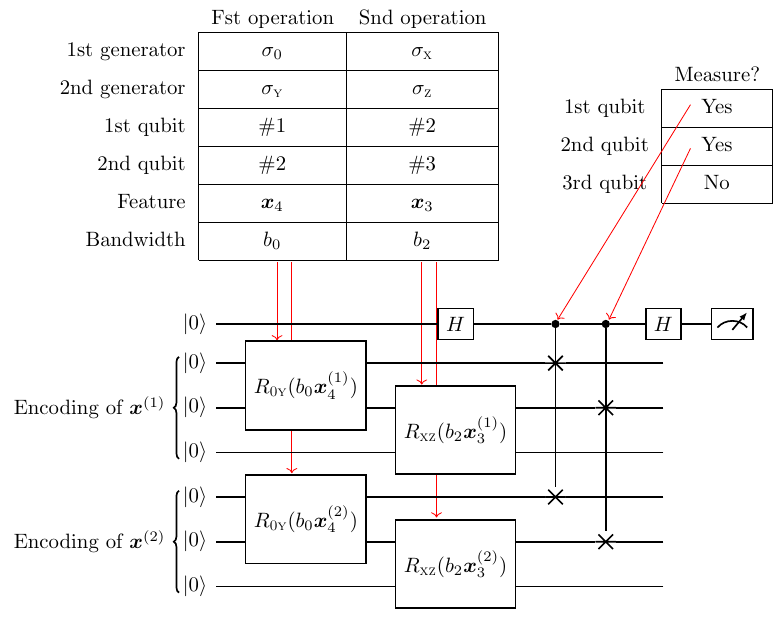}
    \caption{An example of a quantum kernel constructed using our approach. This quantum kernel is characterized by a feature map operating on a Hilbert space with $n = 3$ qubits and consists of $m = 3$ operations. Each sample is represented as $\bm{x} \in \RR^5$ (only the last two features are effectively used in this example). While the bandwidth values are not specified, a common choice is to set $\beta_i = i/10$ for $i \in 1, ..., 10$. The measurement is performed on a subset of the three qubits, making this a projected quantum kernel.}
    \label{fig:circuit_modeling}
\end{figure}

An example of a quantum kernel constructed with our method is shown in Figure~\ref{fig:circuit_modeling}. Each object is described by a vector $\bm{k} \in \mathcal{QK}_{n,m} \subseteq \mathbb{N}^{6m+n}$, and the search space has dimensionality
\begin{equation}
    (4 \times 4 \times n \times (n-1) \times d \times b)^m \times 2^n.
\end{equation}
In fact, for each operation, there are $4 \times 4 \times n \times (n-1) \times d \times b$ possible values, corresponding to each combination of four generators, $n$ possible qubits for the first generator, $n-1$ for the second, $d$ features, and $b$ bandwidth values. Equation 15 follows by considering $m$ operations and all possible qubit measurements over $n$ qubits, the latter resulting in $2^n$ possible values.

\subsection{Expressivity control}\label{sec:expressibility_control}

An important aspect in the design of quantum kernels is to ensure that the quantum transformation defining the kernel is not \emph{too expressive}. In fact, as the recent literature suggests, expressive quantum transformations may lead to ineffective machine learning models. For example, \textcite{huang2021projected} shows that highly expressive unitaries generate full-rank kernel matrices, while it is common to aim for low-rank kernel matrices, which usually lead to simpler, more efficient models and, in many practical settings, are used to approximate kernel matrix. Also, \textcite{kubler2021inductive} analyze kernels in terms of their spectral bias \cite{canatar2021spectral}, focusing on the eigendecomposition of the kernel function (cf. Equation~\ref{eq:mercer}), where components with small eigenvalues imply a large amount of data to be learned, and in the extreme case, components with zero eigenvalues are considered out-of-RKHS, i.e., not learnable. Finally, \textcite{thanasilp2022exponential} demonstrate that highly expressive kernels suffer from the exponential concentration of kernel values, specifically, $\mathrm{Var}[\kappa(\bm{x}, \bm{x}')] \in O(2^{-n})$ with $n$ representing the number of qubits, and $\mathrm{Var}$ is the variance. Consequently, an exponential number of shots is required to distinguish between kernel values. \red{The high-rank kernel matrices, flat eigenvalue distribution, and exponential concentration of kernel values are three distinct phenomena connected by the same root cause: the high dimensionality associated with mapping feature vectors \emph{uniformly} over the exponentially sized Hilbert space, c.f. \textcite{spigler2020asymptotic}.}

For these reasons, we have taken great care in considering expressivity control techniques in our formulation. Specifically, we use the bandwidth parameter $b$, which takes a different value for each gate. As shown in \cite{canatar2022bandwidth}, by multiplying the feature by a small prescaling factor restricts the mapping of elements to a multidimensional cone within the Hilbert space. This approach allows us to retain the beneficial statistical properties of quantum kernels, including having some large eigenvalues in Mercer's eigendecomposition as described in Equation~\ref{eq:mercer}. 

The other type of expressivity control we consider in our method is known as the \emph{projected quantum kernel} (\textcite{huang2021projected}), defines a kernel function as follows:
\begin{equation}\label{eq:projected_quantum_kernel}
\kappa(\bm{x}^{(1)}, \bm{x}^{(2)}) = \optrace[\tilde{\rho}_{\bm{x}^{(1)}} \tilde{\rho}_{\bm{x}^{(2)}}],
\end{equation}
where $\tilde{\rho}_{\bm{x}} = \optrace_Q[U^\dagger(\bm{x}) \rho_0 U(\bm{x})]$ is the reduced density matrix obtained from the $n$-qubit state encoding $\bm{x}$, followed by a partial trace operation to estimate a subsystem \red{of the qubits in $Q \subset \{1, \ldots, n\}$, with $|Q| \ll n$}. Our `partial inner product' approach, as implemented in Figure~\ref{fig:circuit_cswap_test} by applying the swap test gate to a subset of the $n$ target qubits of each state, effectively realizes the function described in Equation~\ref{eq:projected_quantum_kernel}.

\subsection{Criteria}

At the beginning of this section, we have highlighted the significance of establishing criteria for evaluating a quantum kernel $\bm{k} \in \mathcal{QK}_{n, m}$. These criteria serve a dual purpose: they can provide guidance during optimization, as shown in Equation \ref{eq:generic_cost_function}, and can also be employed to analyze candidate solutions, either confirming their effectiveness or revealing their shortcomings.

We have identified three critical aspects when dealing with quantum kernels. The first aspect is \emph{efficient classical simulability}, where transformations achievable on classical hardware, while not necessarily useless, are not significant in the context of seeking quantum computing advantages. The second aspect is \emph{expressivity}, as highly expressive unitaries can lead to unfavorable statistical properties and should be avoided (cf Section~\ref{sec:expressibility_control}). The third aspect is \emph{kernel compatibility}, which involves quantifying how well the chosen kernel aligns with the specific task at hand. It's worth noting that designing hard-to-simulate, mildly expressive quantum kernels that do not work for certain tasks (\emph{no free lunch theorem}) is entirely possible. In fact, many quantum kernels do not perform as well as classical ones in various tasks. Even when they show improvement, figuring out why a particular quantum circuit produces such results can be challenging. An automated procedure can help clarify why a specific quantum kernel excels, indicating that the results meet, to some extent, all the criteria we consider essential for a useful quantum kernel. Here, we introduce several useful criteria, explaining their purposes and quantifying their computational cost.

\red{These criteria can be used to both guide the optimization and debug the solution obtained. Whenever we obtain solutions that are `too simple' i.e., made only with single-qubit gates, we can modify the criteria to take into consideration the rank of the DLA, penalizing solutions having a rank below a certain threshold, and thus avoiding classically simulable solutions. In contrast, solutions that are too expressive might be diagnosed by the low value of the task-model alignment resulting from a flat eigenvalue distribution. Therefore, we can include these criteria within the cost function to avoid this unwanted behavior. The choice of the criteria has to be made by trial and error.}

\subsubsection{Norm of the super-operator $A$} 

The norm of the super-operator $A$ defined in Equation~\ref{eq:norm_A} defines the expressivity of a quantum transformation $U(\cdot)$ and can also represent the expressivity of a quantum kernel. When explicitly constructed, this operator has dimension $2^{n^t} \times 2^{n^t}$. Even for $t = 2$ and modest values of $n$, computing it is infeasible in general.

\subsubsection{Rank of the dynamical Lie algebra} 

The dynamical Lie Algebra (DLA) associated with the quantum transformation $U(\bm{x}) = \prod_{g = 1}^m R_{\alpha_g, \beta_g}^{(q_g, q_g')}(b_{j_g} \bm{x}_{k_g})$ forms a vector space over $\CC$ equipped with a bilinear operation, the commutator, denoted by $[\sigma, \eta] = \sigma\eta - \eta\sigma$. This vector space is spanned by generators of $U$, which include any $\sigma_{\alpha_g}^{(q_g)} \sigma_{\beta_g}^{(q_g')}$ as well as generators derived through repeated commutation of these elements. \textcite{larocca2022diagnosing} have established that a unitary with an associated DLA of dimensionality $O(\exp(n))$ is highly expressive. This criterion is a coarse proxy for expressivity since, when constraining the rotational angles of individual gates, as done when using a bandwidth, it is possible to end up with a unitary that is only mildly expressive but possesses an exponentially sized DLA. \red{Note that in the limit case of bandwidth approaching zero, we obtain minimal expressivity}. Nonetheless, this measure proves extremely valuable, not only for quantifying expressivity but also for assessing the computational complexity of classical simulation. In fact, as demonstrated in \cite{somma2006efficient}, a quantum computation with a polynomially-sized DLA can be efficiently simulated on a classical device. Our choice to work with generators, rather than explicit gates like CNOT, is motivated by the ease of handling the associated DLA in our approach.

From an algorithmic perspective, it remains an open problem to determine the rank of a DLA in time polynomial in $n$, in general. The iterative approach, which involves repeatedly commuting previously identified generators until reaching a fixed point, becomes exponential in $n$ as there can be $4^{n-1}$ possible generators. However, we can approximate this number by establishing an arbitrary threshold $T$ and classifying any unitary as `hard to classically simulate' if its DLA surpasses this rank threshold.

\subsubsection{(Centered) Kernel-target alignment} 

The Kernel-Target alignment (KTA), introduced by \textcite{cristianini2001kernel}, aims to quantify how well a kernel function fits a specific task. Let $\{(\bm{x}^{(j)}, y^{(j)}) \}_{j=1}^p$ be the training set. The KTA is calculated as the (normalized) Frobenius inner product between the kernel Gram matrix $K = [\kappa(\bm{x}^{(i)}, \bm{x}^{(j)})]_{i,j=1}^p$ and the matrix $Y = \bm{y} \bm{y}^\top$, where $\bm{y} = [y_i]_{j=1}^p$ is the vector of labels,
\begin{equation}\label{eq:kta}
\mathrm{A}(K, Y) = \frac{\langle K, Y \rangle_F}{\sqrt{\lVert K \rVert_F \lVert Y \rVert_F}}.
\end{equation}
\textcite{cortes2012algorithms} later enhanced the metric by introducing the \emph{centered} KTA, which subtracts the mean value from both matrices. Their experiments have demonstrated that this modified metric exhibits a stronger correlation with the generalization error across most tasks. Moreover, this approach is a computationally inexpensive and pragmatically effective method for quantifying kernel compatibility.

\subsubsection{Task-model alignment}

The task-model alignment, introduced by \textcite{canatar2021spectral}, offers an alternative approach to quantifying kernel compatibility. It is based on the concept of \emph{spectral bias}, which suggests that when a kernel function $\kappa$ is decomposed into its real, non-negative eigenvalues $\{ \lambda_j \}_{j=0}^\infty$ (sorted in non-decreasing order) and corresponding eigenfunctions $\{ \phi_j \}_{j=0}^\infty$, components associated with larger eigenvalues are easier to learn, requiring fewer samples. Thus, if we eigendecompose the target function of our task as
\begin{equation}
    f(\bm{x}) = \sum_{j = 0}^\infty w_j \sqrt{\lambda_j} \phi_j(\bm{x}),
\end{equation}
we can observe that when $f$ has large $w_j$ associated with its first components, it indicates that the target function can be learned more easily. This concept gives rise to the task-model alignment, which quantifies how much of the \emph{power} of the target function lies within the top $k$ kernel eigenfunctions, 
\begin{equation}
    T(k) = \frac{\sum_{j = 0}^k \lambda_j w_j^2}{\sum_{j = 0}^\infty \lambda_j w_j^2}.
\end{equation}
It's important to note that the spectral bias can be used to explain why a specific kernel function performs well or poorly by examining the contributions of the individual components. If it is not possible to carry out the eigendecomposition analytically, a numerical approach can be used, using the kernel Gram matrix $K$ and the labels $\bm{y}$. \red{It is shown in  \cite[Methods]{canatar2021spectral} that the task-model alignment does not require the knowledge of the target function itself but can be calculated from the training data.}

\subsubsection{Validation error}

The most direct method to assess the effectiveness of a quantum kernel is by computing the accuracy of the classifier concerning a validation test of i.i.d. sampled data points and labels. While this approach can be effective, it is not free from drawbacks: it requires the training of the associated model, introducing overhead that might become too heavy for the task. Additionally, it does not provide insights into the classifier's performance beyond identifying which subset of samples it fails to classify correctly.

\subsection{Optimization}

The optimization technique is used to solve Equation~\ref{eq:optimization}. Due to the non-convexity of the cost function, we can only use heuristic optimization techniques, providing us with no guarantees on the optimality of the solution; \red{as such, it could potentially fail due to the exponentially increasing dimensionality of the search space, inefficacy of the optimization criteria selected, and convergence to suboptimal solutions}. Our approach is independent of any specific optimization technique. In our software, we have implemented several of them. A more detailed description of these techniques is given in 
Supplementary Materials S-IV.

\subsubsection{Greedy approach} This approach iterates over each `cell' of the kernel $\bm{k} \in \mathcal{QK}_{n, m}$ and selects the best value for that single element, then repeats the operation for the following elements. Each action is locally optimal. The process is deterministic once fixed the initial solution. 

\subsubsection{Genetic algorithms} This approach begins with a population of randomly generated candidate solutions. It then proceeds through a three-step process: selection, crossover, and mutation. The selection process extracts a subset of solutions for processing in each iteration, typically favoring those with the best cost function values. The crossover operation combines two candidate solutions, creating, for instance, a solution with the first half of the elements from one parent and the second half from another. Finally, the mutation phase perturbs one or more components of the candidate solution. All resulting solutions are added to the population, and the process repeats. The specific methods used in each of these steps give rise to a wide variety of genetic algorithms. For a comparison of these methods, we refer the reader to \cite{katoch2021review}.

\subsubsection{Bayesian optimization} 

This approach relies on two components: a probabilistic model of the target function and an \emph{acquisition function}. The probabilistic model, typically a Gaussian process, describes the distribution of candidate estimator functions based on the observations $(\bm{x}_i, f(\bm{x}_i))$ and the model's uncertainty. With the addition of more observations, the distribution of functions becomes narrower and closer to the true target. The acquisition function samples the next point to observe. This choice often involves a tradeoff between regions with high variance, where sampling can help reduce uncertainty and improve the model (\emph{exploration}), and regions closer to a previously found optimal value, which aids in converging toward a desirable candidate solution (\emph{exploitation}).

\subsubsection{Reinforcement learning (RL)}

This approach relies on an agent that explores the environment, making a sequence of decisions to maximize a \emph{cumulative} reward function. At each step, the agent's set of possible actions involves adding another operation to the quantum circuit. In this case, the agent can choose the generator, pair of qubits, feature, and bandwidth that characterize that operation. Upon completing the process, the algorithm has determined a \emph{policy}, which establishes a mapping between the states of the algorithm and the actions to be taken. Among the numerous possible Reinforcement Learning (RL) approaches, we have \red{made available} SARSA (State-Action-Reward-State-Action). For an in-depth description, we refer the reader to \cite{matsuo2022deep}.

\section{Application to Anomaly Detection in High Energy Physics}\label{sec:application}

The benchmark application we selected is a kernel-based \emph{anomaly detection} task in HEP. Specifically, the search for new physics processes in Large Hadron Collider (LHC) events with a realistic dataset eventually deployed at CERN in the current scientific campaign. The interest within the HEP community and beyond for these kinds of searches is continuously growing, and one of the clear and vital examples is shown by \textcite{wozniak2023quantum}, where the anomaly search can significantly benefit from the application of quantum kernel techniques, making it a notable and important application of QML.

Our experiments aim to determine whether our automated approach can match or surpass the quantum kernels proposed in \cite{wozniak2023quantum}, which currently represent one of the state-of-the-art this task together with ~\cite{schuhmacher2023unravelling} where the authors find that employing a Quantum Support Vector Classifier trained to identify the artificial anomalies allows to identify realistic BSM events with high accuracy. \red{We clarify that these works are state-of-the-art in regard to kernel methods, and the authors have shown that their quantum kernel outperforms both classical and previously defined quantum kernels, while a more general statement regarding the other possible machine learning model has not been investigated yet.}
To this end, we maintained the same experimental setup and machine learning model used in their study, with the sole modification being the quantum kernel. Our results showcase the capability to achieve performance that matches or surpasses the state-of-the-art results for any process examined. Importantly, these results were obtained without the need to manually design the quantum transformation, underscoring the inherent automation in our approach. A more detailed background on this application can be found in 
Supplementary Material~S-III. 

\subsection{Dataset}\label{sec:dataset}

The dataset used for this study consists of proton-proton collisions simulated at a center-of-mass energy of $\sqrt{13} TeV$. The dataset comprises a combination of the SM process and the BSM processes. These processes are representative of potential new physics scenarios at the Large Hadron Collider (LHC) experiments and are considered anomalies. It is worth noting that we are comparing our results with suboptimal classical methods that do not represent the current state-of-the-art for the community. This is primarily due to the significant feature reduction we applied to the original dataset to accommodate the current constraint on qubit count.

The datasets used in this work are publicly available at \cite{maurizio_pierini_2023_7673769} and represent dijet collision events. Each jet contains 100 particles, and its dynamics according to an SM or a BSM process is simulated via Monte Carlo-based techniques. The resulting data is processed to emulate the effect of the detectors. The BSM processes considered are the Randall-Sundrum gravitons decaying into W-bosons (narrow G), the broad Randall-Sundrum gravitons decaying into W-bosons (broad G), and a scalar boson A decaying into Higgs and Z bosons (A\textrightarrow{}HZ). The simulation results in a large number of features, which are then processed via an autoencoder to a smaller but arbitrary number of features. More details on this construction are shown in 
Supplementary Material~S-III. 
The output dimension of the autoencoder is the \emph{latent dimension} $\ell = 4$ and $\ell = 8$.

After the work of the autoencoder, each of these datasets takes the form $\{ (\bm{x}^{(i)}, y^{(i)}) \in \RR^{2\ell} \times \{\textsc{sm}, \textsc{bsm}\}\}_{i=1}^p$. The factor of $2\ell$ arises from the fact that we are studying \emph{dijet events}, where two jets collide with each other. As a result, we have $\ell$ features for each jet. The features are scaled to the range $(-1, 1)$. Note that, to allow a fair comparison between our results and the ones in \cite{wozniak2023quantum}, the datasets are used as-is. The quantum model is then trained only on Standard Model data and tested to recognize anomalies in unseen data.

\subsection{Setup}

Our experimental assessment has two distinct phases. In the first phase, the \emph{kernel discovery}, we select the kernel $\kappa$, which is represented numerically as $\bm{k} \in \mathcal{QK}$, using our automated approach. In the second phase, the \emph{kernel assessment}, we evaluate the performance of the chosen candidate solution $\bm{k}$ and compare it with the state-of-the-art performance of other kernel methods in the existing literature. We have considered datasets with latent dimensions $\ell = 8$.

During the first phase, our methods are applied to the datasets described in Section \ref{sec:dataset}. We optimize based on a single criterion, accuracy over a (small) validation set. The cost function is defined as 
\begin{equation}\label{eq:cost_hep}
    C(\bm{k}) = \frac{1}{|\mathcal{D}'|} \sum_{(\bm{x}, y) \in \mathcal{D}'} \mathds{1}(f_{\bm{k}}(\bm{x}) = y)
\end{equation}
In this formula, $f_{\bm{k}}$ represents a one-class SVM (OC-SVM) trained on a subset of 75 i.i.d. samples and evaluated on a \emph{different} subset $\mathcal{D}'$ of 75 i.i.d. samples. The function $\mathds{1}(\cdot)$ equals 1 when the predicate specified in its argument is true and 0 otherwise.

Bayesian optimization is employed as the optimization technique, well-suited for complex black-box optimization processes like the one at hand. We start with an initial kernel, which corresponds to the identity matrix. During the kernel discovery phase, we conducted tests using various optimization approach configurations. Our quantum circuit uses a number of qubits equal to the latent dimension, while the number of operations was set to either $m = 8$ or $m = 12$. \red{These values were chosen to balance the trade-off between having a more expressive and complex function and the ease of the optimization process, considering the exponential growth of the dimensionality of the search space and the potential decrease of fidelity on NISQ devices with a higher number of operations.} Since Bayesian optimization is an iterative process, we experimented with 5 and 10 iterations. In each iteration, we sampled 5 points via the acquisition function. Furthermore, we have tested other configurations of our tools, comprising several optimizer and criteria, which are detailed in 
Supplementary Material~S-V.

In the second phase, we evaluate the candidate solution $\bm{k}$ by training the selected model, once again an OC-SVM, and assessing its performance using testing accuracy. We use separate training and testing sets, consisting of 200 and 1500 elements, respectively. The testing process is repeated five times, each with a different i.i.d. samples testing set of 1500 elements, enabling us to estimate the average and standard deviation of accuracy.

It is important to note the difference in data size between the first and second phases. In the second phase, the training process is executed only once, which enables us to analyze a significantly larger dataset within a reasonable timeframe. Conversely, the optimization process is iterative and computationally expensive, requiring careful design of a feasible cost function. Finally, the quantum circuits have been simulated in a noiseless simulation environment using \emph{quask} \cite{dimarcantonio2023quantum} and Qiskit libraries in Python 3.

\subsection{Results}

\begin{table}[tbp]
    \centering
    \scalebox{0.75}{\begin{tabular}{lcccc}
    \toprule
             & Dataset & Dataset & Dataset \\
    Approach & Narrow G & A $\rightarrow$ HZ & Broad G \\
    \midrule
    Best classical kernel (in \cite{wozniak2023quantum}) & $99.54 \pm 0.42$ & $97.94 \pm 0.69$ & $50.90 \pm 3.97$ \\
    Best quantum kernel (in \cite{wozniak2023quantum}) & $99.65 \pm 0.23$ & $98.05 \pm 0.58$ & $55.20 \pm 3.96$ \\
    This work  ($m=8,t= 5$) & $99.28 \pm 0.58$ & $97.66 \pm 0.61$ & $47.62 \pm 4.17$ \\
    This work  ($m=8,t=10$) & $99.28 \pm 0.58$ & $97.66 \pm 0.61$ & $\textbf{61.61} \pm \textbf{4.67}$ \\
    This work  ($m=12,t= 5$) & $99.37\pm 0.46$ & $94.57\pm 1.60$ & $58.70\pm 4.38$ \\
    This work  ($m=12,t=10$) & $\textbf{99.73} \pm \textbf{0.20}$ & $\textbf{98.34} \pm \textbf{0.37}$ & $58.70 \pm 4.38$ \\
    \bottomrule
    \end{tabular}}\smallskip
    \caption{Comparison of the top testing accuracy (average and standard deviation) of the best classical and quantum kernels for proton collision anomaly detection, with a latent dimension $\ell = 8$. For quantum approaches, the number of qubits is equal to $\ell$. The best quantum kernel in \cite{wozniak2023quantum} has of $2\ell$ rotational gates and $\ell - 1$ \textsc{cnot}. The best quantum kernel with our approach has $m$ two-qubit rotational gates, $m = 8, 12$. Bayesian optimization has been run for $t = 5, 10$ iterations. The best-performing approach is highlighted in bold.}
    \label{tab:results_l8}
\end{table}

We compare our approach to the best performing classical kernel and the state-of-the-art quantum kernel  in \cite{wozniak2023quantum}, focusing on a latent dimension  $\ell = 8$. In our analysis, we examine the performance of the final model, measured as the Area Under the Curve (AUC) of the Receiver Operating Characteristic (ROC) curve, by varying the parameters of interest $m$, representing the number of operations (which can be $8$ or $12$), and $t$, denoting the number of optimization epochs (either $5$ or $10$). 

For the Narrow G and A$\rightarrow$ HZ datasets (see Table~\ref{tab:results_l8}), our method with $m=12$ operations outperforms both classical and quantum kernels. By using a reduced number of operations ($m=8$), our algorithm discovers a kernel with a slightly lower accuracy than the state-of-the-art quantum kernel, suggesting that a more complex circuit is required to better capture and leverage the underlying patterns in the data. Nevertheless, both datasets have nearly reached a point of saturation, providing minimal opportunity for further enhancement. On the other hand,  for the Broad G dataset, which is notably more challenging to analyze, our algorithm discovers a kernel able to outperform both classical and quantum kernels by a consistent margin. It is interesting to note that on this dataset the maximum performance is achieved on $m=8$, suggesting that having a smaller number of gates simplifies the optimization process, leading to a better overall performance. This is in agreement with the consideration that Bayesian optimization, which has been chosen for its efficacy in black-box optimization processes, is known to face challenges when dealing with problems that involve a high number of variables. Bridging this observation with our results, we note that with $m=12$ we still obtain a better AUC than both compared methods, despite the increased complexity.

\begin{figure}[tbp]
    \centering
    \includegraphics[width=0.45\linewidth]{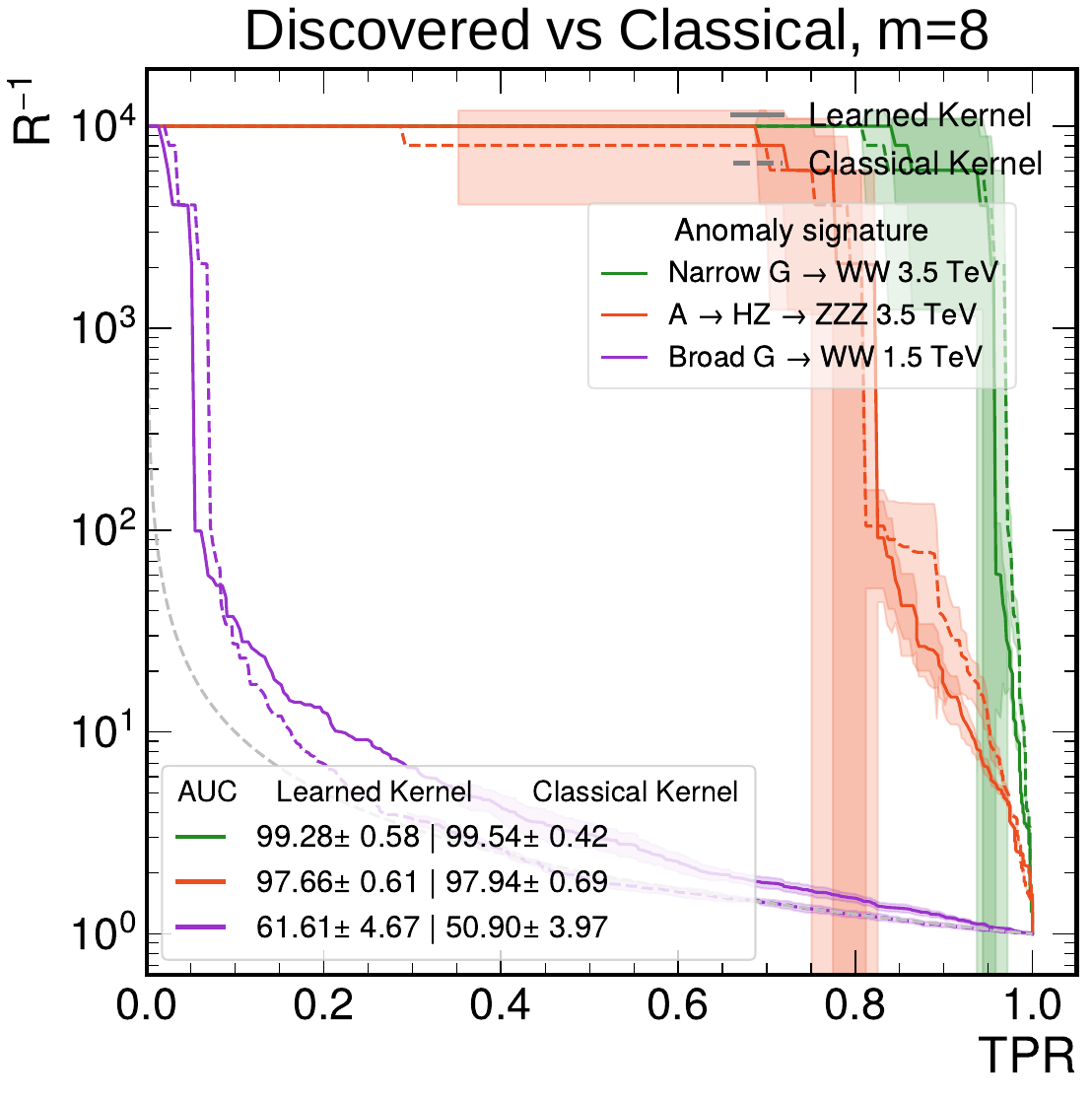}~
    \includegraphics[width=0.45\linewidth]{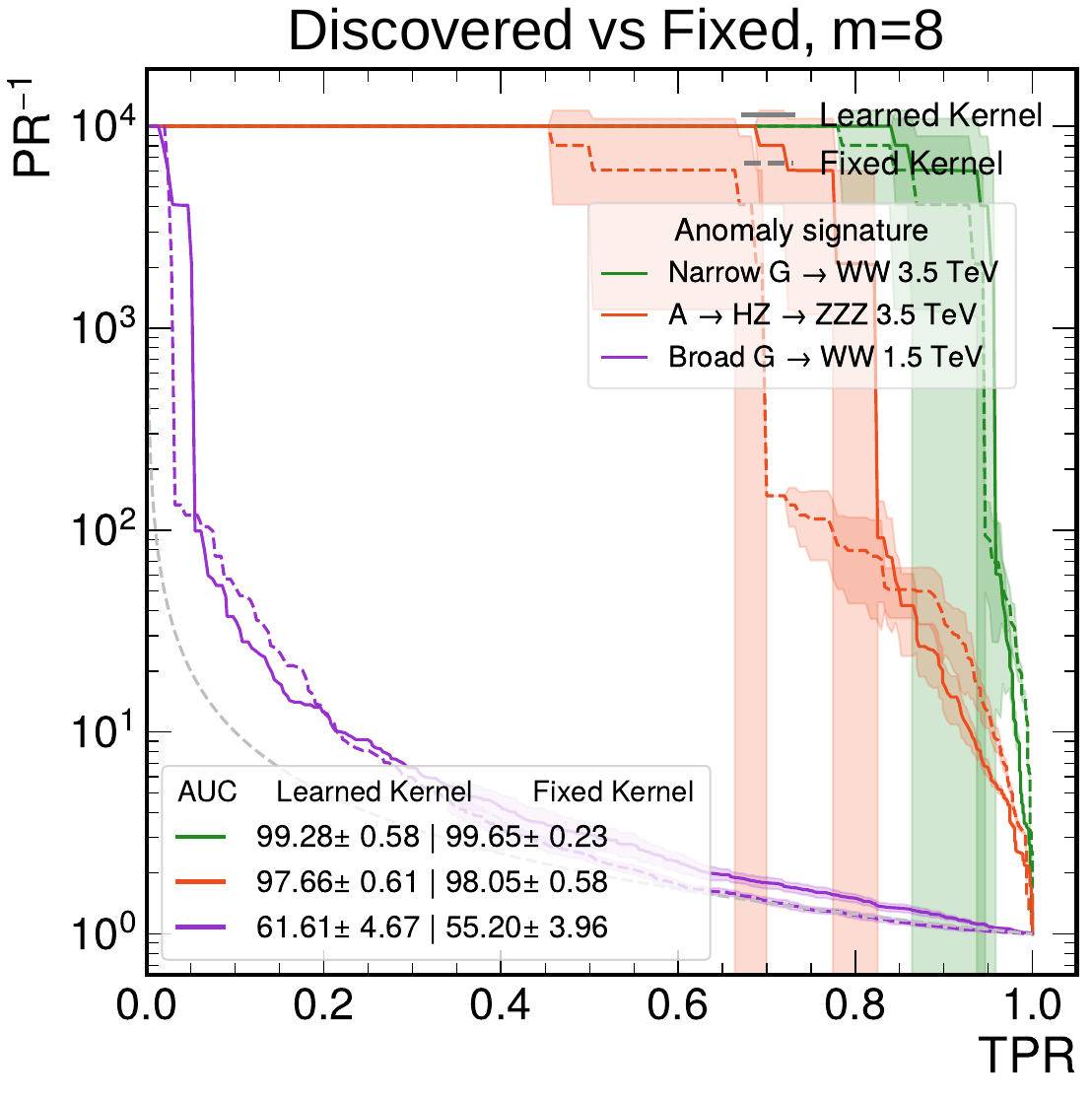}
    \caption{ROC-AUC curve comparing our approach (solid line) with the best classical (dashed line, left) and quantum (dashed line, right) kernels from the literature for datasets with a latent dimension of 8. \red{The shaded areas represent a standard deviation of 1 from the average value (solid line).} In the displayed configuration, the kernel built with our approach consists of $m=8$ operations.}
    \label{fig:roc_latent8_gate8}
\end{figure}

\begin{figure}[tbp]
    \centering
    \includegraphics[width=0.45\linewidth]{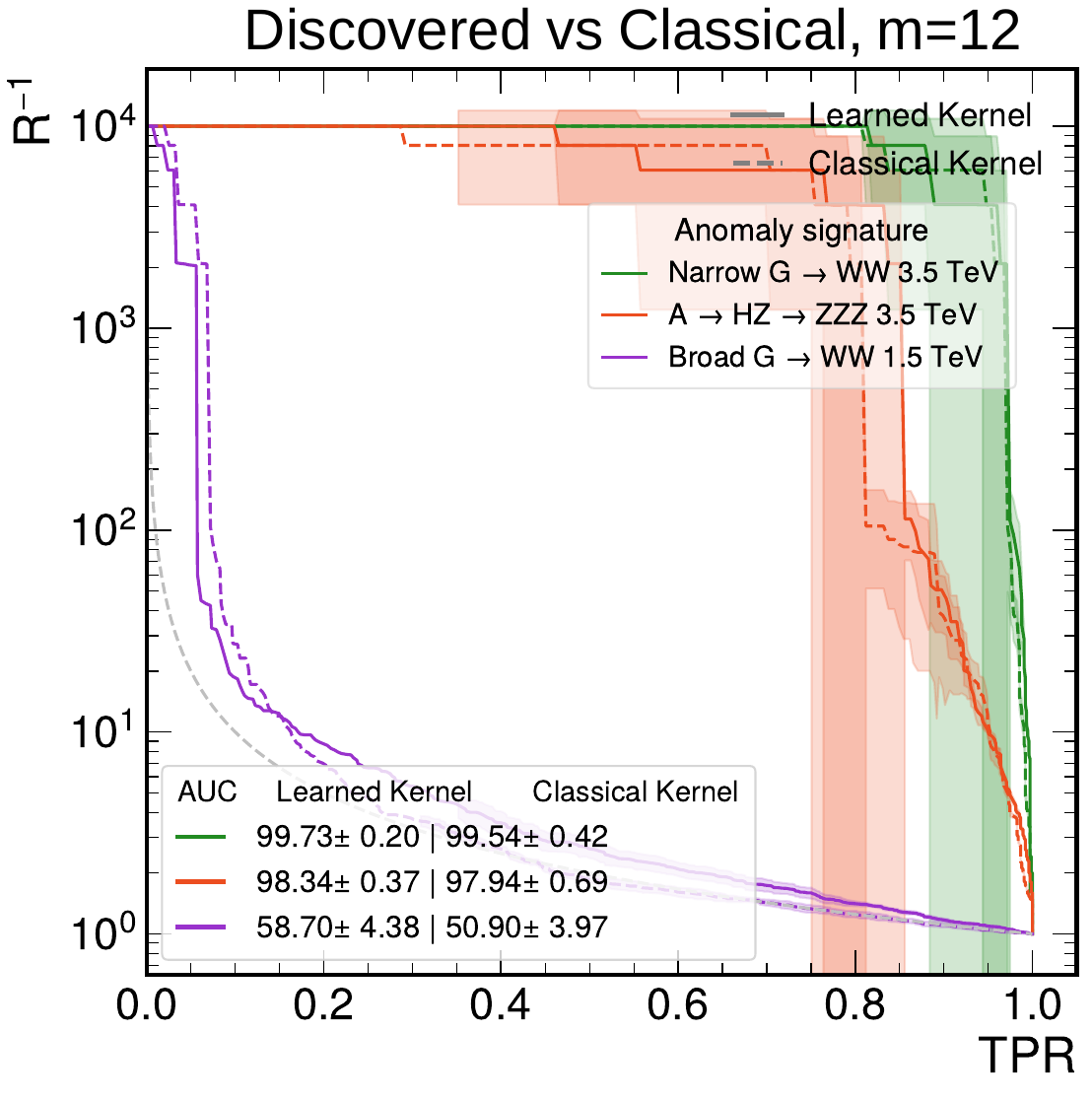}~
    \includegraphics[width=0.45\linewidth]{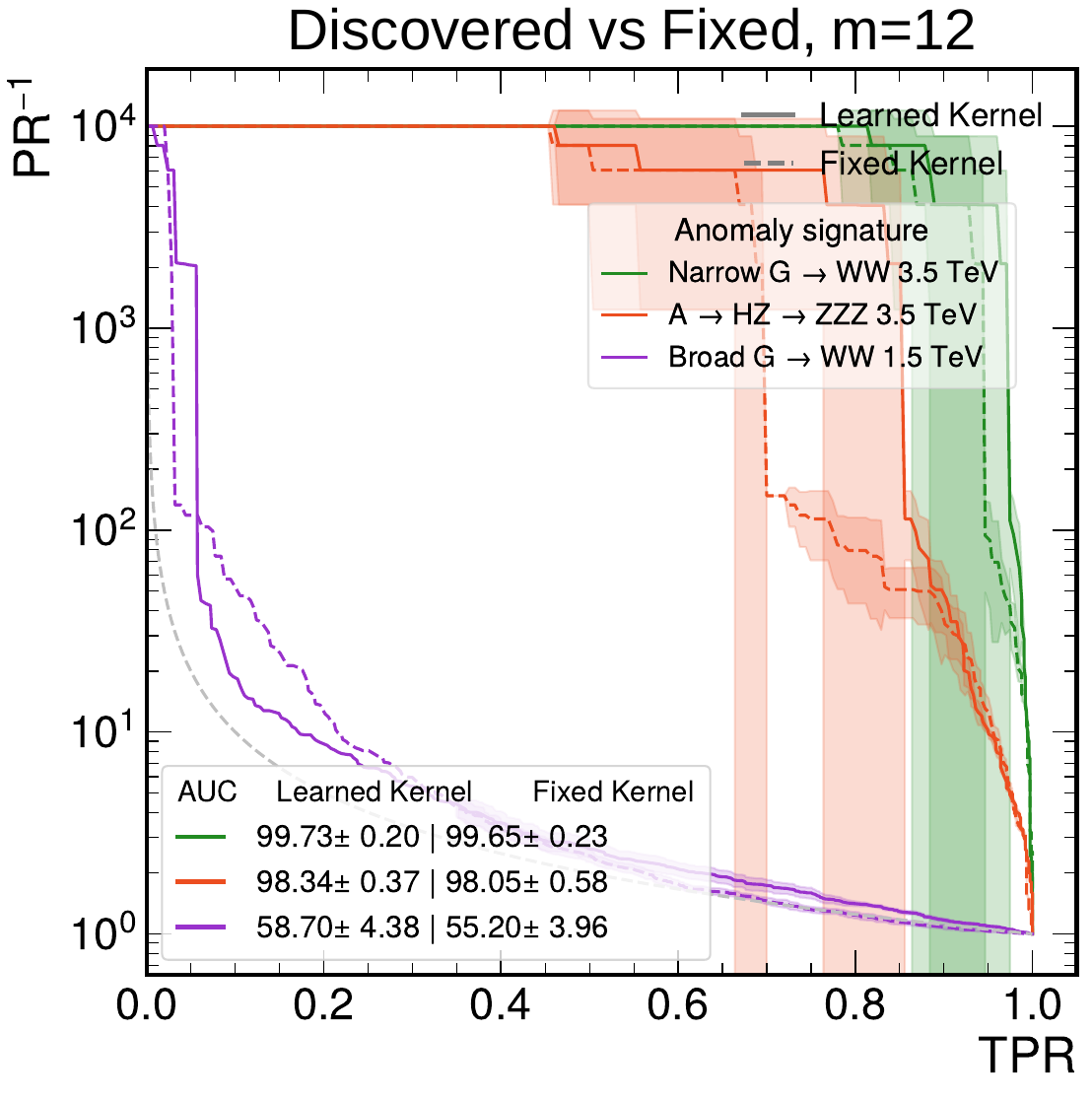}
    \caption{ROC-AUC curve comparing our approach (solid line) with the best classical (dashed line, left) and quantum (dashed line, right) kernels from the literature for datasets with a latent dimension of 8. \red{The shaded areas represent a standard deviation of 1 from the average value (solid line).} In the displayed configuration, the kernel built with our approach consists of $m=12$ operations.}
    \label{fig:roc_latent8_gate12}
\end{figure}

A more comprehensive performance analysis is given by considering Receiver Operating Characteristic (ROC) curves (see Figures ~\ref{fig:roc_latent8_gate8} and ~\ref{fig:roc_latent8_gate12}). These curves illustrate the relationship between the True Positive Rate (TPR), which signifies correctly classified events, and the False Positive Rate (FPR), representing misclassified events, while varying the threshold parameter $\nu$ of the model, that acts as an upper bound on the fraction of outliers (see 
Supplementary Material~S-II
for more details).  Specifically, we highlight the configurations that benefit the most from our approach with a latent dimension of $8$, considering circuits with $8$ operations (see Figure~\ref{fig:roc_latent8_gate8}) and $12$ operations (see Figure~\ref{fig:roc_latent8_gate12}). We compare these results to the ROC of the best classical and best quantum kernels. 
By considering the ROC, we can examine for which and how many values of $\nu$ a particular model is more accurate than another. For example, as shown in Figure~\ref{fig:roc_latent8_gate8} for the Broad G dataset, our discovered kernel with $8$ operations exhibits a lower FPR (and therefore a better accuracy) than the state-of-the-art quantum kernel for each value of TPR lower than 0.075, and for each value between 0.2 and 1. In general, every discovered model with a higher AUC shows a lower FPR for the vast majority of TPR values, demonstrating an overall superior performance. \red{The results for the other configurations of optimizers and criteria are detailed in 
Supplementary Material~S-V
and are comparable to the results here obtained with Bayesian optimization and validation accuracy as criteria.}

\section{Conclusions}\label{sec:conclusion}

We have presented a novel approach to address the challenging task of selecting an appropriate quantum kernel for a specific task. We have introduced an algorithm for the automated discovery of quantum kernels, which operates without prior knowledge of the underlying problem, relying solely on a set of observations, and yields a candidate solution through heuristic optimization. We demonstrate the effectiveness of our approach in solving the problem of proton collision anomaly detection, a crucial issue in high-energy physics and a significant benchmark in the field of quantum machine learning due to its potential for quantum advantage. In the tested configuration, our approach identifies quantum kernels that either match or surpass the performance of the best-performing methods currently known. We have made our software available as open-source, enabling researchers to explore its potential benefits in other use cases.

While the proposed approach offers a broad range of functionality and configuration options, encompassing various criteria and optimization techniques, we have only presented here a limited subset of the possible configurations. We have left for future research the investigation of the individual impacts of each criterion and a comparison of the results obtained with different optimization algorithms. Additionally, an interesting path to improve our method could be to select optimization algorithms well-suited for a large number of variables, in order to explore potential solutions with a higher number of operations.  This could even involve variants of Bayesian optimization, such as Sparse Axis-Aligned Subspaces (SAAS-BO), which impose sparsity assumptions on the variables \cite{eriksson2021high}.

Moreover, a promising direction for further exploration lies in identifying new tasks with data that share a similar nature to proton collision anomaly detection. These tasks could prove to be excellent candidates for practical applications of quantum kernels.

\section{Data and Code Availability}

The datasets are available at \cite{maurizio_pierini_2023_7673769}. The software is based on \emph{quask} library for Python 3 \cite{dimarcantonio2023quantum}. The code is accessible at: \url{https://github.com/Dan-LB/Quantum_Kernel_Discovery_AD_HEP}.

\section{Acknowledgements}

MI and ADP are supported by the Istituto Nazionale di Alta Matematica ``Francesco Severi''. MG is supported by CERN through the CERN Quantum Technology Initiative. We acknowledge the CINECA award under the ISCRA initiative.

\printbibliography

\vfill

\newcounter{supsection}
\setcounter{supsection}{1}
\renewcommand{\thesection}{S-\Roman{supsection}}

\clearpage
\onecolumn 

\begin{flushleft}
\makeatletter
\noindent{\Huge{}Supplementary material: \@title}
\makeatother
\end{flushleft}

\begin{refsection}
\section{Related works}\label{sec:relatedworks}

Our work is inspired by numerous machine learning techniques that aim to customize models for specific tasks by optimizing their parameters and configurations. Listing all the relevant literature is beyond the scope of this paper. Therefore, in this section, we explore the analogies between our approach and those previously proposed in classical and quantum machine learning, referring the reader to more comprehensive treatments of the subject when needed.

\subsection{Classical machine learning}

In classical machine learning, an approach closely related to our technique is the \emph{neural architecture search} (NAS), which is aimed at automating the creation of artificial neural networks. This process involves defining a search space to select the best network topology from various possibilities. NAS proves particularly valuable when determining the optimal number of layers and neurons per layer is not readily calculable or guessable. Drawing an analogy between the network underlying a multilayer perceptron and the network underlying a quantum circuit provides strong motivation for applying these techniques in quantum machine learning. Various strategies have been employed to explore the NAS search space, including reinforcement learning \cite{zoph2016neural}, random search \cite{liu2017hierarchical}, and Bayesian optimization \cite{white2021bananas}. NAS is part of a larger family of approaches known as \emph{AutoML}, which focuses on automating the design of a machine learning model across all its stages, including feature preprocessing and selection. While NAS is certainly restricted to artificial neural networks, AutoML comprehends any model, including kernels. Our algorithm, which determines the subset of features to be included in the quantum circuit, falls within the broader category of AutoML. For a comprehensive review of this field, we refer the reader to \textcite{waring2020automated}.

Another task similar to the one at hand is the \emph{symbolic regression}. This technique involves regression methods based on symbolic or mathematical expressions and has found applications in rediscovering physical laws \cite{udrescu2020ai}. In this context, it is common to use a fitness function to minimize both the empirical error and the model complexity.
To address the challenge of combinatorial explosion, they impose a constraint requiring the resulting formula to have meaningful physical units associated with it. Once again, the analogy with our approach lies in the treatment of the basic operations used to construct the symbolic formula, similar to how we handle single operations applied via the quantum circuit. The technique to limit the model complexity is similar to the expressivity control techniques in quantum kernels.
For additional approaches to scientific discovery using artificial intelligence, we recommend referring to the review by \textcite{krenn2022scientific}.

\subsection{Quantum machine learning}

In quantum machine learning, the task of customizing a quantum circuit for a specific task has been widely studied. Two main approaches are typically followed: the variational approach and the combinatorial approach.

The variational approach involves incorporating certain free parameters as rotational arguments within the quantum circuit and training these parameters to minimize a loss function using stochastic gradient descent. Authors in \cite{lloyd2020quantum} have defined a quantum kernel that takes into account, along with the features, some free parameters to be optimized to maximize the Kernel Alignment. These techniques have been later applied to work on data with group structure by \cite{glick2021covariant}. The advantage in this case is the use of a computationally cheap optimization technique and taking advantage of the large body of literature and software available on variational quantum algorithm. Clearly, when using these approaches, we must carefully avoid any cause of barren plateau, including high expressivity.

The combinatorial approach is centered around selecting the sequence of gates that constitute the quantum kernel. The discrete nature of this description requires the use of combinatorial optimization techniques. These methods have been notably useful in the context of variational quantum classifiers. For instance, \cite{ostaszewski2021reinforcement} applied reinforcement learning agents to optimize the quantum circuit for VQE, specifically for learning the ground state of lithium hydride. \cite{meng2021quantum} advocated the use of Monte Carlo Tree Search as an effective heuristic for circuit optimization, while \cite{huang2022robust} demonstrated the efficacy of evolutionary algorithms. Additionally, \cite{du2022quantum} proposed a technique to optimize both the ansatz and the parameters of the variational quantum circuit in a single step, minimizing resource usage.

The combinatorial approach applied to quantum kernels has not been studied in such depth. \textcite{altares2021automatic} pioneered the use of genetic algorithms to construct the feature map of a quantum kernel. Their feature map is structured similarly to our approach, representing quantum gates ($\textsc{cnot}$, $R_\textsc{x}$, $R_\textsc{y}$, $R_\textsc{z}$) instead of using the generators as in our approach. The authors do not explicitly consider the expressivity of the model but provide a measure of complexity in terms of the number of single-qubit gates and the number of CNOT gates separately. The drawback of this approach is that penalizing CNOT gates results in quantum circuits that can be efficiently simulated on classical devices. A similar approach has been tested in \cite{chen2022genetically} using synthetic datasets. None of these approaches explicitly use theoretical results in the theory of quantum kernels. 

\setcounter{supsection}{2}
\section{Derivation of the one-class SVM}\label{sec:oneclasssvm}

SVM is typically employed to address binary classification problems, although it can be adapted to tackle various tasks, including anomaly detection. Anomaly detection is often achieved using a \emph{one-class SVM} (OC-SVM), an unsupervised learning model that identifies samples that appear to be drawn from a different distribution than the regular one. This model is capable of handling both novelty detection and outlier detection. The distinction between these tasks lies in the training data; novelty detection is trained exclusively on `regular' data, while outlier detection is trained on both regular data and anomalies. The task addressed in this work, the anomaly detection in proton collision events, falls under the category of novelty detection.

In this section, we show the functioning of such a model. Our explanation first introduce the hard- and soft-margin SVMs, which is commonly used in classification problems, then adapt the formulation to suit the unsupervised nature of novelty detection.

\subsection{Hard-margin SVM}

Consider a binary classification task. Let $\{(\bm{x}^{(j)}, y^{(j)})\}_{j=1}^p$ be the training set, where $y^{(j)} \in \{\pm 1\}$. Suppose the dataset is linearly separable, meaning there exists a hyperplane $\bm{w}^\top \bm{x} + b = 0$, with $\bm{w}$ as the normal vector to the hyperplane and $b$ as the offset to the origin, which perfectly separates the two classes.

An SVM finds a pair of parallel hyperplanes such that one class's samples lie on one side of the first hyperplane and the other class's samples are in the opposite region of the second hyperplane, and the distance between the two hyperplanes is maximized. This problem can be modeled as a constrained optimization:
\begin{align}
& \arg\min_{\bm{w},b} \langle \bm{w}, \bm{w} \rangle \\ 
& \text{subject to } y^{(j)}(\bm{w}^\top \bm{x}^{(j)} + b) \ge 1, \text{for }j = 1, \ldots, p
\end{align}
The solution leads to hyperplanes that are distant by $2/\lVert \bm{w}\rVert$ from each other. The resulting classifier, known as the \emph{hard-margin SVM}, has the following prediction function:
\begin{equation}
    \tilde{f}(\bm{x}) = \mathrm{sign}(\bm{w}^\top \bm{x} + b).
\end{equation}

We can solve the optimization problem using Lagrangian multiplier methods, introducing a slack variable $\alpha_j$ for each constraint, leading to:
\begin{equation}
    \mathcal{L}(\bm{w}, \bm{\alpha}) = \frac{1}{2} \langle w, w \rangle - \sum_{j=1}^p \bm{\alpha}_j [y^j (\bm{x}^j \bm{w} + b) - 1].
\end{equation}
The problem can be solved in its primal form $\arg\min_{\bm{w}, b} \max_{\alpha} \mathcal{L}(\bm{w}, \bm{\alpha})$ or in its dual form $\arg \max_{\alpha} \min_{\bm{w}, b} \mathcal{L}(\bm{w}, \bm{\alpha})$. For this particular problem, the primal and dual forms lead to equivalent solutions. We prefer to solve the dual problem because it immediately works with kernel functions. We first set $\partial \mathcal{L}/\partial \bm{w} = \partial \mathcal{L}/\partial b = 0$, which leads to the optimization problem:
\begin{equation}
    \arg\max_{\bm{\alpha} > 0} - \frac{1}{2} \sum_{j,k=1}^p \mathbf{\alpha}_j \mathbf{\alpha}_k y^j y^k \langle \mathbf{x}^j, \mathbf{x}^k \rangle 
+ \sum_{j=1}^p \mathbf{\alpha}_j
\end{equation}

This optimization problem is convex and can be solved efficiently using quadratic programming. The solution of the dual Lagrangian leads to a classifier in the form:
\begin{equation}
    \tilde{f}(\bm{x}) = \mathrm{sign}\left(\sum_{j \in SV} \alpha_j \langle \bf{x}, \bf{x}^{(j)} \rangle \right).
\end{equation}
Here, the elements $j \in SV$ are called support vectors. This optimization problem often leads to sparse solutions, as the optimal value for any $\bm{\alpha}_j$ is zero. The non-zero values are associated with samples lying on the hyperplane contributing to the solution. For this reason, inference with an SVM is computationally efficient. Note that the `dual form' of the SVM allows us to use kernel methods without any additional effort by replacing the Euclidean inner product with any $\kappa$. 

\subsection{Soft-margin SVM}

To account for noisy datasets that might not be linearly separable, the \emph{soft-margin SVM} can be used. It relies on a set of $m$ slack variables $\xi_j = \max(0, 1 - y^{(j)}(\bm{w}^\top \bm{x}^{(j)} + b))$, representing the classification error for the $j$-th element of the training set. The optimization problem is then formulated as follows:
\begin{align}
& \arg\min_{\bm{w},b} \langle \bm{w}, \bm{w} \rangle + C \sum_{j=1}^p \xi_j \\ 
& \text{subject to } y^{(j)}(\bm{w}^\top \bm{x}^{(j)} + b) \ge 1 - \xi_j, \text{for }j = 1, \ldots, m.
\end{align}
The constant $C \in [0, +\infty]$ penalizes these errors, and for $C = \infty$, we recover the hard-margin SVM. 

Alternatively, we can formulate the same model differently using the $\nu$-SVM (nu-SVM), replacing the parameter $C$ with $\nu \in [0,1)$. Note that $\nu$ is an upper bound on the fraction of margin errors and a lower bound on the fraction of support vectors. The formulation is as follows:
\begin{align}
& \arg\min_{\bm{w}, b, \xi \ge 0, \rho>0} \frac{1}{2} \langle \bm{w}, \bm{w} \rangle - \nu\rho + \frac{1}{2} \sum_{j=1}^p \xi_j \\ 
& \text{subject to } y^{(j)}(\bm{w}^\top \bm{x}^{(j)} + b) \ge \rho - \xi_j, \text{for } j = 1, \ldots, m.
\end{align}

It's worth noting that $\rho$ substitutes $1$ in the constraint, indicating that the margins are now at a distance of $2\rho/\lVert \bm{w} \rVert$, not $2/\lVert \bm{w} \rVert$ as in the hard-margin case.

\subsection{One-class SVM}

Finally, we introduce the one-class SVM \cite{scholkopf1999support}. The optimization problem is defined as follows:
\begin{align}
    & \arg\min_{w, b, \xi \ge 0, \rho \neq 0} \frac{1}{2} \langle w, w \rangle - \rho + \frac{1}{\nu d} \sum_{j=1}^p \xi_j \\ 
    & \text{subject to } (\mathbf{w}^\top \mathbf{x}^j + b) \ge \rho - \xi_j \text{ for } j = 1, \ldots, m.
\end{align}
Here, $d$ is the dimensionality of the feature vectors. It's important to note that in this formulation, we do not consider labels, which allows us to work with unsupervised tasks. 

The dual form of the Lagrangian associated with this task is as follows:
\begin{align}
    & \arg\min_{0 \le \bm{\alpha}_j \le \frac{1}{\nu d}, \sum_j \bm{\alpha}_j = 1} \frac{1}{2} \sum_{j,k=1}^p \bm{\alpha}_j \bm{\alpha}_k \langle \bm{x}^{(j)}, \bm{x}^{(k)} \rangle
\end{align}
The solution to this leads to the function:
\begin{equation}
    \tilde{f}(\bm{x}) = \mathrm{sign}\left(\sum_{j \in SV} \alpha_j \langle \bf{x}, \bf{x}^{(j)} \rangle - \rho\right).
\end{equation}
For $\tilde{f}(\bm{x}) = 1$, the data is inferred to be regular, while for $\tilde{f}(\bm{x}) = -1$, it is inferred to be anomalous. Again, the Euclidean inner product can be replaced with any kernel function.

\setcounter{supsection}{3}
\section{Extended background on anomaly detection in proton collision}\label{apx:proton}

\subsection{Motivation}

After the discovery of the Higgs boson, one of the main goals of the LHC physics program is the search for new phenomena that would answer some of the open questions associated with the Standard Model (SM) of particle physics, the theory that describes our understanding of the fundamental constituents of the universe such as the origin of Dark Matter and Dark Energy and why the known forces in nature are characterized by largely different energy scales.
The scientific workflow for the community consists, on the one hand, of collecting experimental data from proton beams that are accelerated in the collider and, on the other hand, of generating precise simulated data with Montecarlo techniques. Those collected and generated data are then compared to depict deviations with respect to the adopted theoretical formulation of Nature, namely the SM. 

In this setting, we define searches for Beyond Standard Model (BSM) physics-model-dependent approach where a signal is defined as a specific process described by the chosen BSM scenario, and the background is defined as any SM process that generates a similar signature as that of a BSM process. 
Anomaly detection searches can be described then as a model-independent approach, where there is no need to postulate a priori the signal of interest to enable a more sensitive analysis in case an unexpected signature is present in the data. Consequently, this unsupervised approach considers anomalies as any event in the data that deviates from the SM predictions relying minimally on specific new physics scenarios. 
In this work, we look specifically, as done in similar work, to new exotic particles decaying to jet pairs, namely, a resonance (peak) in the dijet mass $(m_{jj})$ distribution. 

\subsection{Detailed construction of the dataset}

The dataset consists of Monte Carlo simulated samples of standard model physics with the addition of BSM processes related to the actual sensitivity of LHC experiments. 
To allow a realistic loading and analysis of the adopted dataset on current quantum devices the data are mapped to a latent representation of reduced dimensionality using an autoencoder model.

The datasets used in this work are publicly available at \cite{maurizio_pierini_2023_7673769}, it represents a dijet collision event. Each jet contains 100 particles and each particle is described by three spatial coordinates for a total of 300 features. 
The BSM processes (aka anomalies) to benchmark the model performance are: Randall-Sundrum gravitons decaying into W-bosons (narrow G), broad Randall-Sundrum gravitons decaying into W-bosons (broad G), and a scalar boson A decaying into Higgs and Z bosons (A\textrightarrow{}HZ). 
Events are produced with PYTHIA~\cite{pythia}, Jets are clustered from reconstructed particles using the anti-kT~\cite{antikt} clustering algorithm, with typical kinematic cut emulating the effect of a typical LHC online event selection. Subsequently, data are processed with DELPHES~\cite{delphes} to emulate detector effects, specifically with the adoption of CMS configuration card. 

Then, this huge dataset is manipulated using a classical autoencoder that is trained to compress particle jet objects individually without access to truth labels. This approach is preferred over the more standard PCA method due to its ability to capture nonlinear relationships within the data. The output dimension of the autoencoder is the \emph{latent dimension}. The original work in \cite{wozniak2023quantum} has considered multiple possible latent dimensions, while here we consider only $\ell = 4$ and $\ell = 8$.

We have used the pre-processed dataset, after the simulation and the autoencoding. These datasets takes the form $\{ (\bm{x}^{(i)}, y^{(i)}) \in \RR^{2\ell} \times \{\textsc{sm}, \textsc{bsm}\}\}_{i=1}^p$. The factor of $2\ell$ arises from the fact that we are studying \emph{dijet events}, where two jets collide with each other. As a result, we have $\ell$ features for each jet. 

\setcounter{supsection}{4}
\section{Extended description and runtime of the optimization algorithm}\label{apx:optimization}

Here we more extensively detail the description of the optimzation techniques implemented within our software. 

\subsection{Greedy optimization}

The greedy optimization method is our simplest heuristics. Our implementation starts from the initial candidate provided as input and proceeds by iterating through all the operations indexed from $i = 1$ to $m$. During each iteration, we change the first cell of the candidate solution and select the one that minimizes the cost function. While this perturbation occurs, the other cells remain unchanged. Following this, we proceed to fix the first cell and sequentially perturb and fix the subsequent cells to their locally optimal values. 

This approach is particularly useful during the exploration phase, as its moderate computational cost allows to explore multiple configurations. From an implementation perspective, the greedy optimizer has been written from scratch and does not rely on any dependency other than the plain Python3 library. The runtime is $(4+4+n+(n-1)+d+b) \times m$ where $n$ number of qubits, $m$ number of operations, $d$ number of features, $b$ number of bandwidth values. This runtime has to be multiplied by the cost of evaluating the candidate solution, which ultimately depends on the criteria chosen.

\subsection{Genetic optimization}

The genetic optimization is a population-based approach inspired by natural selection and genetics. A moltitude of variants of genetic optimizers exists. In our case, we have relied on the version developed within the library \texttt{opytimizer} \cite{de2019opytimizer} for Python3. The algorithms works as follows:
\begin{enumerate}
    \item We define a vector of variables that will be the object of the optimization, each component is an integer with lower and upper bound fixed a priori. Each component corresponds to a cell of a certain operation within the quantum kernel.
    \item The algorithm begins by initializing a population of potential solutions, named agents, generated randomly. Each agent represents a candidate solution to the optimization problem. The population size and the representation of solutions are configurable parameters.
    \item In the \emph{selection} phase, agents are chosen from the population for reproduction based on their fitness (inverse of cost function). Fitness is determined by evaluating each agent's solution using the objective function of the optimization problem. Agents with higher fitness values are more likely to be selected for reproduction. The default choice for the fraction of agent selected is 0.75.
    \item Selected agents, referred to as parents, undergo \emph{crossover} to produce new candidate solution, named children. The crossover operation introduces diversity into the population by combining different genetic material (cell values) from successful solutions. The probability of crossover determines how often this operation occurs, which is by default 0.5.
    \item Children resulting from crossover may undergo \emph{mutation}, where random changes are introduced to their genetic information. Mutation helps explore new regions of the search space and prevents the algorithm from getting stuck in local optima. The mutation probability controls the likelihood of mutation occurring in each offspring, which is by default 0.25.
    \item After crossover and mutation, the fitness of each offspring is evaluated using the objective function. This step determines how well each solution performs with respect to the optimization criteria. The children, along with the parent population, form the next generation. Survivor selection ensures that the population size remains constant throughout the algorithm's execution. Various survivor selection strategies can be employed, in this case we use the Roulette selection. It assigns a probability of selection to each candidate solution based on its fitness relative to the total fitness of the population.
    \item These steps are iterated an arbitrary number of times, at the end the solution with the highest fitness is returned as solution of the optimization problem.
\end{enumerate}

The runtime is $O(p \times i)$, $p$ is the population size and $i$ the number of iterations. The cost of generation, crossover, mutation is negligible compared to the cost of the fitness function. This runtime has to be multiplied by the cost of evaluating the candidate solution, which ultimately depends on the criteria chosen.

\subsection{Bayesian optimization}

Bayesian optimization is a heuristic optimization algorithm ideally suited for optimizing black-box functions with highly expensive evaluations and a relatively low number of variables. Our implementation uses the \texttt{scikit-optimize} library for Python3. The algorithm operates as follows:

\begin{enumerate}
    \item We initialize an initial set of $4$ randomly generated points $x_i, i = 1, ...4$ and the corresponding observations $f(x_i)$, $f$ function to optimize which in our case is the criteria evaluating the quantum kernel.
    \item For each iteration \( t = 1, ..., T \):
    \begin{enumerate}
        \item we add four new observations to the set \( \{ (x_i, f(x_i)) \}_{i=1}^{4(t+1)} \) and construct a probabilistic model for the function \( f \) based on these observations. The integration over all possible functions within the probabilistic model serves as our temporary candidate solution, achieved through Gaussian process regression.
        \item To determine the next point \( x_{t+1} \) to sample and generate the next observation, we optimize a computationally inexpensive 'acquisition' function or surrogate model. In our case, the acquisition function is the Probability of Improvement (PI), 
        \begin{equation}
            \text{PI}(x) = P(f(x) \leq f_{\text{best}} - \xi)
        \end{equation}
        where $P$ is the cumulative distribution function, $f(x)$ is the predicted value at $x$, $f_\text{best}$ is the best observed value of the objective function so far, $\xi$ is a parameter controlling the trade-off between exploration and exploitation.
    \end{enumerate}
    \item Finally, after \( T \) iterations, the integration over the probabilistic model provides us with the best modeling for our function, allowing us to efficiently locate its minimum value.
\end{enumerate}

The fitting of observation within the Gaussian Regression Process has runtime $O(n^3)$, $n$ number of sampled points. The optimization of $PI(x)$ require itself solving another optimization problem.

\subsection{SARSA optimization}

For our reinforcement learning algorithm, we utilized the \texttt{Mushroom-RL} library for Python3 \cite{d2021mushroomrl}, specifically employing its \texttt{SARSALambda} implementation of the SARSA algorithm. While providing a comprehensive explanation of reinforcement learning and SARSA($\lambda$) is beyond the scope of this article, interested readers can find detailed information in \cite{matsuo2022deep}. 

\setcounter{supsection}{5}
\section{Extended experimental evaluation}\label{apx:experiments}

The experiment presented in Section~\ref{sec:application} is limited by the arbitrary choice of the optimizer, which has been fixed to Bayesian optimization, and the criteria, arbitrarily set to an accuracy measure of the SVM over a validation test. While this initial configuration is reasonable, it may not necessarily yield the best solutions. Bayesian optimization is selected for its adaptability in exploring costly black-box functions, where the computational expense of the Gaussian Process and surrogate model is negligible compared to the cost of computing the criteria. This approach often produces decent results with few iterations. The validation accuracy is chosen as the simplest and most interpretable criterion.

However, potentially better solutions may be obtained by exploring other optimizers and criteria. While attempting all possible configurations offered by our tool is impractical and beyond the scope of this paper, we provide a comparison with alternative approaches for the sake of completeness. Specifically, we compare the results obtained with Bayesian optimization to those from Reinforcement Learning-based optimization, and the validation accuracy to a linear combination (equal weights) of the validation accuracy and the task-model alignment. These configurations are evaluated across all three datasets.

\begin{figure}[p]
    \centering
    \includegraphics[width=0.9\linewidth]{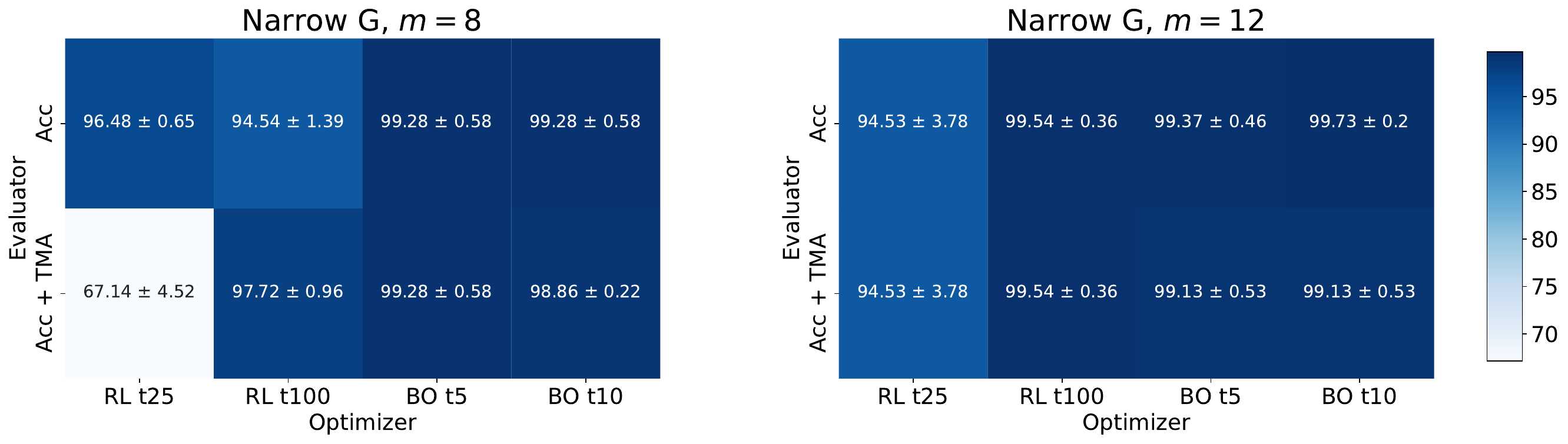}
    \caption{Heatmap showing the results, computed as the Area under the Curve of the ROC and considered with the corresponding uncertainty, on the Narrow G dataset. Experiments with the Bayesian Optimizer (labeled as BO) are performed as described in Section \ref{sec:application}. Experiments with SARSA (RL) are conducted for $25$ and $100$ episodes. The number of operations $m$ for the kernels are respectively $8$ on the left and $12$ on the right.}
    \label{fig:extended_task_1}
\end{figure}

In Figure~\ref{fig:extended_task_1}, we tested the Narrow G dataset. Both optimizers behaved similarly, with the RL engine yielding worse solutions when trained for fewer than $25$ epochs. Except for one negative result, using either criterion for optimization resulted in an accuracy change from $1\%$ to $3\%$.

\begin{figure}[p]
    \centering
    \includegraphics[width=0.9\linewidth]{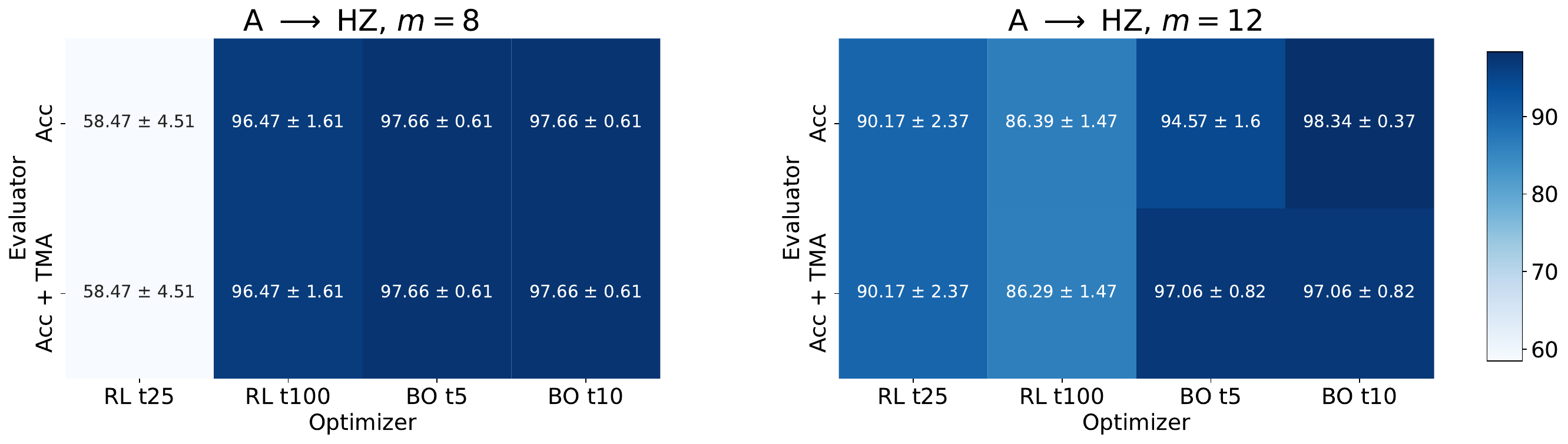}
    \caption{Heatmap showing the results, computed as the Area under the Curve of the ROC and considered with the corresponding uncertainty, on the A$\rightarrow$HZ dataset. Experiments with the Bayesian Optimizer (labeled as BO) are performed as described in Section \ref{sec:application}. Experiments with SARSA (RL) are conducted for $25$ and $100$ episodes. The number of operations $m$ for the kernels are respectively $8$ on the left and $12$ on the right.}
    \label{fig:extended_task_2}
\end{figure}

In Figure~\ref{fig:extended_task_2}, we tested the  A $\rightarrow$ HZ dataset. Similar behavior to the previous dataset was observed: the optimizers were comparable when the RL engine was trained for a sufficient number of epochs. However, depending on the case, changing the criterion could lead to small improvements.

\begin{figure}[p]
    \centering
    \includegraphics[width=0.9\linewidth]{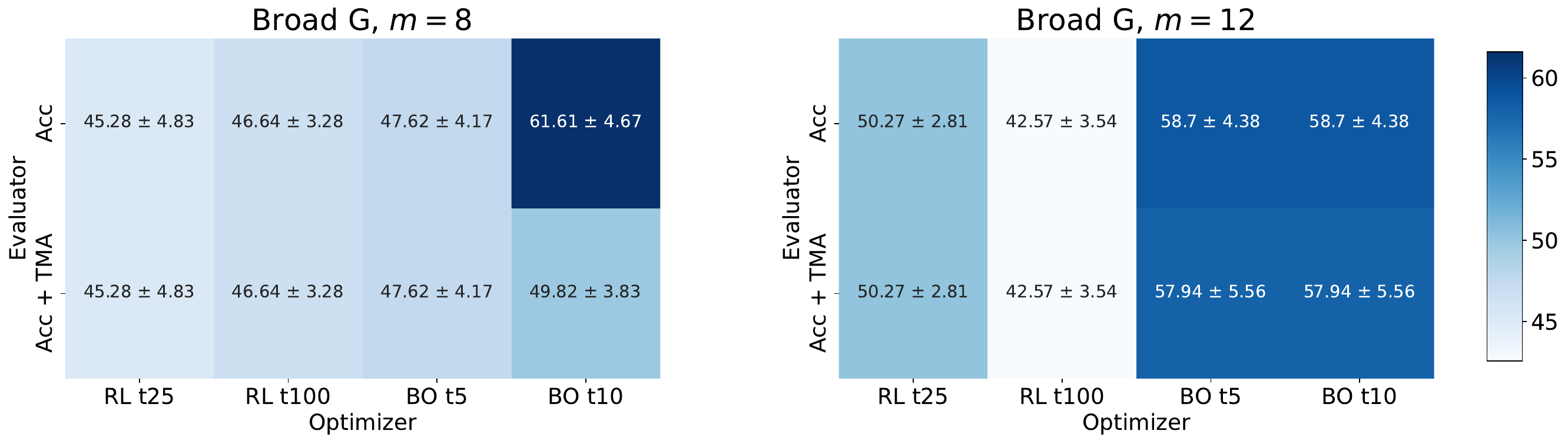}
    \caption{Heatmap showing the results, computed as the Area under the Curve of the ROC and considered with the corresponding uncertainty, on the Broad G dataset. Experiments with the Bayesian Optimizer (labeled as BO) are performed as described in Section \ref{sec:application}. Experiments with SARSA (RL) are conducted for $25$ and $100$ episodes. The number of operations $m$ for the kernels are respectively $8$ on the left and $12$ on the right.}
    \label{fig:extended_task_3}
\end{figure}

In Figure~\ref{fig:extended_task_3}, we tested the Broad G dataset. The configuration had no effect on the results for a number of gates $m = 12$, while for smaller numbers of gates, the configuration using BO and validation accuracy showed substantial improvement over the others. Notably, RL trained for more epochs could sometimes yield worse results, induced by the stochasticity of the optimization process.

We conclude with a brief comment on the different criteria, providing detailed guidelines on how to choose them and the expected behavior of including a particular criterion within the cost function to optimize. The norm of the super operator precisely quantifies the expressibility of the quantum circuit; however, it serves primarily as a diagnostic tool for a candidate solution over a fairly small number of qubits due to its extremely high computational cost. It should primarily be used to investigate a kernel that does not generalize well, as an excess of expressibility may be mitigated by employing expressibility control techniques.
The rank of the Dynamical Lie Algebra serves as both a coarse-grained proxy for expressibility (an exponentially sized DLA without any expressibility control implies the unitary is highly expressible) and as a tool to assess the efficiency of classical simulability (a too-small DLA is classically simulable). It should be incorporated into the optimization process as it is not overly costly to calculate, especially if we limit ourselves to checking the DLA up to a certain threshold. This criterion can be used to discard unitaries that are too simple to allow for a quantum advantage.
Both the centered Kernel-Target alignment (KTA) and the Task-Model alignment (TMA) measure the compatibility between the task and the chosen kernel. The former is easier to calculate and can be readily added to the optimization process, while the latter is excellent for diagnosing the kernel's effectiveness and can also be used in the optimization phase, albeit at a higher computational cost than the KTA. Finally, the validation accuracy truly evaluates the performance of the kernel function on a model, providing an approximation of its generalization error. While more precise than KTA and TMA, it is also much more expensive. It should primarily be used to evaluate the quality of a candidate solution at the end of the optimization or within the optimization of fairly small quantum kernels.

\printbibliography[title=Supplementary material references]

@book{steinwart2008svmbook,
	title        = {Support Vector Machines},
	author       = {Steinwart, Ingo and Christmann, Andreas},
	year         = 2008,
	publisher    = {Springer},
	isbn         = {0387772413}
}

@book{manentimotta2023quantumbook,
	title        = {Quantum Information Science},
	author       = {Manenti, Riccardo and Motta, Mario},
	year         = 2023,
	publisher    = {Oxford University Press},
	isbn         = 9780198787488
}

@article{d2021mushroomrl,
  title={Mushroomrl: Simplifying reinforcement learning research},
  author={D'Eramo, Carlo and Tateo, Davide and Bonarini, Andrea and Restelli, Marcello and Peters, Jan},
  journal={Journal of Machine Learning Research},
  volume={22},
  number={131},
  pages={1--5},
  year={2021}
}

@article{de2019opytimizer,
  title={Opytimizer: A nature-inspired python optimizer},
  author={de Rosa, Gustavo H and Rodrigues, Douglas and Papa, Jo{\~a}o P},
  journal={arXiv preprint arXiv:1912.13002},
  year={2019}
}

@book{schuld2021qmlbook,
	title        = {Machine learning with quantum computers},
	author       = {Schuld, Maria and Petruccione, Francesco},
	year         = 2021,
	publisher    = {Springer},
	isbn         = {978-3-030-83098-4},
	edition      = {2nd}
}

@article{schuld2019qmlfeaturehilbert,
	title        = {Quantum Machine Learning in Feature Hilbert Spaces},
	author       = {Maria Schuld and Nathan Killoran},
	year         = 2019,
	month        = 2,
	journal      = {Physical Review Letters},
	publisher    = {American Physical Society ({APS})},
	volume       = 122,
	number       = 4,
	doi          = {10.1103/physrevlett.122.040504},
	url          = {https://doi.org/10.1103%2Fphysrevlett.122.040504}
}

@article{havlivcek2019supervised,
	title        = {Supervised learning with quantum-enhanced feature spaces},
	author       = {Havlicek, Vojtech and C{\'o}rcoles, Antonio D and Temme, Kristan and Harrow, Aram W and Kandala, Abhinav and Chow, Jerry M and Gambetta, Jay M},
	year         = 2019,
	journal      = {Nature},
	publisher    = {Nature Publishing Group},
	volume       = 567,
	number       = 7747,
	pages        = {209--212}
}

@article{biamonte2017quantum,
	title        = {Quantum machine learning},
	author       = {Biamonte, Jacob and Wittek, Peter and Pancotti, Nicola and Rebentrost, Patrick and Wiebe, Nathan and Lloyd, Seth},
	year         = 2017,
	journal      = {Nature},
	publisher    = {Nature Publishing Group UK London},
	volume       = 549,
	number       = 7671,
	pages        = {195--202}
}

@article{muandet2017kernel,
  title={Kernel mean embedding of distributions: A review and beyond},
  author={Muandet, Krikamol and Fukumizu, Kenji and Sriperumbudur, Bharath and Sch{\"o}lkopf, Bernhard and others},
  journal={Foundations and Trends{\textregistered} in Machine Learning},
  volume={10},
  number={1-2},
  pages={1--141},
  year={2017},
  publisher={Now Publishers, Inc.}
}

@article{liu2021rigorous,
	title        = {A rigorous and robust quantum speed-up in supervised machine learning},
	author       = {Liu, Yunchao and Arunachalam, Srinivasan and Temme, Kristan},
	year         = 2021,
	journal      = {Nature Physics},
	publisher    = {Nature Publishing Group UK London},
	volume       = 17,
	number       = 9,
	pages        = {1013--1017}
}

@INPROCEEDINGS {incudini2023higher,
author = {M. Incudini and F. Martini and A. Di Pierro},
booktitle = {2023 IEEE International Conference on Quantum Computing and Engineering (QCE)},
title = {Higher-Order Topological Kernels via Quantum Computation},
year = {2023},
volume = {},
issn = {},
pages = {621-629},
doi = {10.1109/QCE57702.2023.00076},
url = {https://doi.ieeecomputersociety.org/10.1109/QCE57702.2023.00076},
publisher = {},
address = {},
month = {9}
}

@article{jager2023universal,
	title        = {Universal expressiveness of variational quantum classifiers and quantum kernels for support vector machines},
	author       = {J{\"a}ger, Jonas and Krems, Roman V},
	year         = 2023,
	journal      = {Nature Communications},
	publisher    = {Nature Publishing Group UK London},
	volume       = 14,
	number       = 1,
	pages        = 576
}

@article{preskill2018quantum,
	title        = {Quantum computing in the NISQ era and beyond},
	author       = {Preskill, John},
	year         = 2018,
	journal      = {Quantum},
	publisher    = {Verein zur F{\"o}rderung des Open Access Publizierens in den Quantenwissenschaften},
	volume       = 2,
	pages        = 79
}

@article{somma2006efficient,
	title        = {Efficient solvability of Hamiltonians and limits on the power of some quantum computational models},
	author       = {Somma, Rolando and Barnum, Howard and Ortiz, Gerardo and Knill, Emanuel},
	year         = 2006,
	journal      = {Physical review letters},
	publisher    = {APS},
	volume       = 97,
	number       = 19,
	pages        = 190501
}

@article{cerezo2022challenges,
	title        = {Challenges and opportunities in quantum machine learning},
	author       = {Cerezo, Marco and Verdon, Guillaume and Huang, Hsin-Yuan and Cincio, Lukasz and Coles, Patrick J},
	year         = 2022,
	journal      = {Nature Computational Science},
	publisher    = {Nature Publishing Group US New York},
	volume       = 2,
	number       = 9,
	pages        = {567--576}
}

@article{ostaszewski2021reinforcement,
	title        = {Reinforcement learning for optimization of variational quantum circuit architectures},
	author       = {Ostaszewski, Mateusz and Trenkwalder, Lea M and Masarczyk, Wojciech and Scerri, Eleanor and Dunjko, Vedran},
	year         = 2021,
	journal      = {Advances in Neural Information Processing Systems},
	volume       = 34,
	pages        = {18182--18194}
}

@article{huang2021projected,
	title        = {{Power of data in quantum machine learning}},
	author       = {{Huang}, Hsin-Yuan and {Broughton}, Michael and {Mohseni}, Masoud and {Babbush}, Ryan and {Boixo}, Sergio and {Neven}, Hartmut and {McClean}, Jarrod R.},
	year         = 2021,
	month        = jan,
	journal      = {Nature Communications},
	volume       = 12,
	pages        = 2631,
	doi          = {10.1038/s41467-021-22539-9},
	eid          = 2631
}

@article{kubler2021inductive,
	title        = {The inductive bias of quantum kernels},
	author       = {K{\"u}bler, Jonas and Buchholz, Simon and Sch{\"o}lkopf, Bernhard},
	year         = 2021,
	journal      = {Advances in Neural Information Processing Systems},
	volume       = 34,
	pages        = {12661--12673}
}

@article{canatar2021spectral,
	title        = {Spectral bias and task-model alignment explain generalization in kernel regression and infinitely wide neural networks},
	author       = {Canatar, Abdulkadir and Bordelon, Blake and Pehlevan, Cengiz},
	year         = 2021,
	journal      = {Nature communications},
	publisher    = {Nature Publishing Group UK London},
	volume       = 12,
	number       = 1,
	pages        = 2914
}

@article{thanasilp2022exponential,
  title={Exponential concentration in quantum kernel methods},
  author={Thanasilp, Supanut and Wang, Samson and Cerezo, M and Holmes, Zo{\"e}},
  journal={Nature Communications},
  volume={15},
  number={1},
  pages={5200},
  year={2024},
  publisher={Nature Publishing Group UK London}
}

@article{
canatar2022bandwidth,
title={Bandwidth Enables Generalization in Quantum Kernel Models},
author={Abdulkadir Canatar and Evan Peters and Cengiz Pehlevan and Stefan M. Wild and Ruslan Shaydulin},
journal={Transactions on Machine Learning Research},
issn={2835-8856},
year={2023},
url={https://openreview.net/forum?id=A1N2qp4yAq},
note={}
}

@article{larocca2022diagnosing,
	title        = {Diagnosing barren plateaus with tools from quantum optimal control},
	author       = {Larocca, Martin and Czarnik, Piotr and Sharma, Kunal and Muraleedharan, Gopikrishnan and Coles, Patrick J and Cerezo, Marco},
	year         = 2022,
	journal      = {Quantum},
	publisher    = {Verein zur F{\"o}rderung des Open Access Publizierens in den Quantenwissenschaften},
	volume       = 6,
	pages        = 824
}

@article{cristianini2001kernel,
	title        = {On kernel-target alignment},
	author       = {Cristianini, Nello and Shawe-Taylor, John and Elisseeff, Andre and Kandola, Jaz},
	year         = 2001,
	journal      = {Advances in neural information processing systems},
	volume       = 14
}

@article{cortes2012algorithms,
	title        = {Algorithms for learning kernels based on centered alignment},
	author       = {Cortes, Corinna and Mohri, Mehryar and Rostamizadeh, Afshin},
	year         = 2012,
	journal      = {The Journal of Machine Learning Research},
	publisher    = {JMLR. org},
	volume       = 13,
	pages        = {795--828}
}

@article{katoch2021review,
	title        = {A review on genetic algorithm: past, present, and future},
	author       = {Katoch, Sourabh and Chauhan, Sumit Singh and Kumar, Vijay},
	year         = 2021,
	journal      = {Multimedia tools and applications},
	publisher    = {Springer},
	volume       = 80,
	number       = 5,
	pages        = {8091--8126}
}

@article{matsuo2022deep,
	title        = {Deep learning, reinforcement learning, and world models},
	author       = {Matsuo, Yutaka and LeCun, Yann and Sahani, Maneesh and Precup, Doina and Silver, David and Sugiyama, Masashi and Uchibe, Eiji and Morimoto, Jun},
	year         = 2022,
	journal      = {Neural Networks},
	publisher    = {Elsevier},
	volume       = 152,
	pages        = {267--275}
}

@article{wozniak2023quantum,
	title        = {Quantum anomaly detection in the latent space of proton collision events at the LHC},
	author       = {Wo{\'z}niak, Kinga Anna and Belis, Vasilis and Puljak, Ema and Barkoutsos, Panagiotis and Dissertori, G{\"u}nther and Grossi, Michele and Pierini, Maurizio and Reiter, Florentin and Tavernelli, Ivano and Vallecorsa, Sofia},
	year         = 2023,
	journal      = {arXiv preprint arXiv:2301.10780}
}

@article{dimarcantonio2023quantum,
	title        = {Quantum Advantage Seeker with Kernels (QuASK): a software framework to speed up the research in quantum machine learning},
	author       = {Di Marcantonio, Francesco and Incudini, Massimiliano and Tezza, Davide and Grossi, Michele},
	year         = 2023,
	journal      = {Quantum Machine Intelligence},
	publisher    = {Springer},
	volume       = 5,
	number       = 1,
	pages        = 20
}

@article{sim2019expressibility,
	title        = {Expressibility and entangling capability of parameterized quantum circuits for hybrid quantum-classical algorithms},
	author       = {Sim, Sukin and Johnson, Peter D and Aspuru-Guzik, Alan},
	year         = 2019,
	journal      = {Advanced Quantum Technologies},
	publisher    = {Wiley Online Library},
	volume       = 2,
	number       = 12,
	pages        = 1900070
}

@dataset{maurizio_pierini_2023_7673769,
	title        = {{Dataset for Quantum anomaly detection in the latent space of proton collision events at the LHC}},
	author       = {Maurizio Pierini and Kinga Anna Wozniak},
	year         = 2023,
	month        = feb,
	publisher    = {Zenodo},
	doi          = {10.5281/zenodo.7673769},
	url          = {https://doi.org/10.5281/zenodo.7673769}
}

@inproceedings{eriksson2021high,
	title        = {High-dimensional Bayesian optimization with sparse axis-aligned subspaces},
	author       = {Eriksson, David and Jankowiak, Martin},
	year         = 2021,
	booktitle    = {Uncertainty in Artificial Intelligence},
	pages        = {493--503},
	organization = {PMLR}
}

@article{scholkopf1999support,
	title        = {Support vector method for novelty detection},
	author       = {Sch{\"o}lkopf, Bernhard and Williamson, Robert C and Smola, Alex and Shawe-Taylor, John and Platt, John},
	year         = 1999,
	journal      = {Advances in neural information processing systems},
	volume       = 12
}

@article{nakata2014generating,
	title        = {Generating a state t-design by diagonal quantum circuits},
	author       = {Nakata, Yoshifumi and Koashi, Masato and Murao, Mio},
	year         = 2014,
	journal      = {New Journal of Physics},
	publisher    = {IOP Publishing},
	volume       = 16,
	number       = 5,
	pages        = {053043}
}

@article{altares2021automatic,
	title        = {Automatic design of quantum feature maps},
	author       = {Altares-L{\'o}pez, Sergio and Ribeiro, Angela and Garc{\'\i}a-Ripoll, Juan Jos{\'e}},
	year         = 2021,
	journal      = {Quantum Science and Technology},
	publisher    = {IOP Publishing},
	volume       = 6,
	number       = 4,
	pages        = {045015}
}

@article{spigler2020asymptotic,
  title={Asymptotic learning curves of kernel methods: empirical data versus teacher--student paradigm},
  author={Spigler, Stefano and Geiger, Mario and Wyart, Matthieu},
  journal={Journal of Statistical Mechanics: Theory and Experiment},
  volume={2020},
  number={12},
  pages={124001},
  year={2020},
  publisher={IOP Publishing}
}

@article{gan2023unified,
  title={A Unified Framework for Trace-induced Quantum Kernels},
  author={Gan, Beng Yee and Leykam, Daniel and Thanasilp, Supanut},
  journal={arXiv preprint arXiv:2311.13552},
  year={2023}
}

@article{glick2021covariant,
	title        = {Covariant quantum kernels for data with group structure},
	author       = {Glick, Jennifer R and Gujarati, Tanvi P and Corcoles, Antonio D and Kim, Youngseok and Kandala, Abhinav and Gambetta, Jay M and Temme, Kristan},
	year         = 2021,
	journal      = {arXiv preprint arXiv:2105.03406}
}

@article{lloyd2020quantum,
	title        = {Quantum embeddings for machine learning},
	author       = {Lloyd, Seth and Schuld, Maria and Ijaz, Aroosa and Izaac, Josh and Killoran, Nathan},
	year         = 2020,
	journal      = {arXiv preprint arXiv:2001.03622}
}

@article{chen2022genetically,
	title        = {Genetically auto-generated quantum feature maps},
	author       = {Chen, Bang-Shien and Chern, Jann-Long},
	year         = 2022,
	journal      = {arXiv preprint arXiv:2207.11449}
}

@article{huang2022robust,
	title        = {Robust resource-efficient quantum variational ansatz through an evolutionary algorithm},
	author       = {Huang, Yuhan and Li, Qingyu and Hou, Xiaokai and Wu, Rebing and Yung, Man-Hong and Bayat, Abolfazl and Wang, Xiaoting},
	year         = 2022,
	journal      = {Physical Review A},
	publisher    = {APS},
	volume       = 105,
	number       = 5,
	pages        = {052414}
}

@article{du2022quantum,
	title        = {Quantum circuit architecture search for variational quantum algorithms},
	author       = {Du, Yuxuan and Huang, Tao and You, Shan and Hsieh, Min-Hsiu and Tao, Dacheng},
	year         = 2022,
	journal      = {npj Quantum Information},
	publisher    = {Nature Publishing Group},
	volume       = 8,
	number       = 1,
	pages        = {1--8}
}

@article{meng2021quantum,
	title        = {Quantum Circuit Architecture Optimization for Variational Quantum Eigensolver via Monto Carlo Tree Search},
	author       = {Meng, Fan-Xu and Li, Ze-Tong and Yu, Xu-Tao and Zhang, Zai-Chen},
	year         = 2021,
	journal      = {IEEE Transactions on Quantum Engineering},
	publisher    = {IEEE},
	volume       = 2,
	pages        = {1--10}
}

@article{zoph2016neural,
	title        = {Neural architecture search with reinforcement learning},
	author       = {Zoph, Barret and Le, Quoc V},
	year         = 2016,
	journal      = {arXiv preprint arXiv:1611.01578}
}

@article{liu2017hierarchical,
	title        = {Hierarchical representations for efficient architecture search},
	author       = {Liu, Hanxiao and Simonyan, Karen and Vinyals, Oriol and Fernando, Chrisantha and Kavukcuoglu, Koray},
	year         = 2017,
	journal      = {arXiv preprint arXiv:1711.00436}
}

@inproceedings{white2021bananas,
	title        = {Bananas: Bayesian optimization with neural architectures for neural architecture search},
	author       = {White, Colin and Neiswanger, Willie and Savani, Yash},
	year         = 2021,
	booktitle    = {Proceedings of the AAAI Conference on Artificial Intelligence},
	volume       = 35,
	number       = 12,
	pages        = {10293--10301}
}

@article{waring2020automated,
	title        = {Automated machine learning: Review of the state-of-the-art and opportunities for healthcare},
	author       = {Waring, Jonathan and Lindvall, Charlotta and Umeton, Renato},
	year         = 2020,
	journal      = {Artificial intelligence in medicine},
	publisher    = {Elsevier},
	volume       = 104,
	pages        = 101822
}

@article{udrescu2020ai,
	title        = {AI Feynman 2.0: Pareto-optimal symbolic regression exploiting graph modularity},
	author       = {Udrescu, Silviu-Marian and Tan, Andrew and Feng, Jiahai and Neto, Orisvaldo and Wu, Tailin and Tegmark, Max},
	year         = 2020,
	journal      = {Advances in Neural Information Processing Systems},
	volume       = 33,
	pages        = {4860--4871}
}

@article{krenn2022scientific,
	title        = {On scientific understanding with artificial intelligence},
	author       = {Krenn, Mario and Pollice, Robert and Guo, Si Yue and Aldeghi, Matteo and Cervera-Lierta, Alba and Friederich, Pascal and dos Passos Gomes, Gabriel and H{\"a}se, Florian and Jinich, Adrian and Nigam, AkshatKumar and others},
	year         = 2022,
	journal      = {Nature Reviews Physics},
	publisher    = {Nature Publishing Group UK London},
	volume       = 4,
	number       = 12,
	pages        = {761--769}
}

@article{pythia,
	title        = {{An Introduction to PYTHIA 8.2}},
	author       = {Sj{\"o}strand, Torbjörn and others},
	year         = 2015,
	journal      = {Comput. Phys. Commun.},
	volume       = 191,
	pages        = {159--177},
	doi          = {10.1016/j.cpc.2015.01.024},
	eprint       = {1410.3012},
	archiveprefix = {arXiv},
	primaryclass = {hep-ph},
	reportnumber = {LU-TP-14-36, MCNET-14-22, CERN-PH-TH-2014-190, FERMILAB-PUB-14-316-CD, DESY-14-178, SLAC-PUB-16122},
	slaccitation = {%%CITATION = ARXIV:1410.3012;%%}
}

@article{antikt,
	title        = {{The anti-$k_t$ jet clustering algorithm}},
	author       = {M. Cacciari and others},
	year         = 2008,
	journal      = {JHEP},
	volume       = {04},
	pages        = {063},
	doi          = {10.1088/1126-6708/2008/04/063},
	eprint       = {0802.1189},
	archiveprefix = {arXiv},
	primaryclass = {hep-ph},
	reportnumber = {LPTHE-07-03},
	slaccitation = {%%CITATION = ARXIV:0802.1189;%%}
}

@article{delphes,
	title        = {{DELPHES 3, A modular framework for fast simulation of a generic collider experiment}},
	author       = {J. de Favereau and others},
	year         = 2014,
	journal      = {JHEP},
	volume       = {02},
	pages        = {057},
	doi          = {10.1007/JHEP02(2014)057},
	collaboration = {DELPHES 3},
	eprint       = {1307.6346},
	archiveprefix = {arXiv},
	primaryclass = {hep-ex},
	slaccitation = {%%CITATION = ARXIV:1307.6346;%%}
}

@article{schuhmacher2023unravelling,
  title={Unravelling physics beyond the standard model with classical and quantum anomaly detection},
  author={Schuhmacher, Julian and Boggia, Laura and Belis, Vasilis and Puljak, Ema and Grossi, Michele and Pierini, Maurizio and Vallecorsa, Sofia and Tacchino, Francesco and Barkoutsos, Panagiotis and Tavernelli, Ivano},
  journal={Machine Learning: Science and Technology},
  volume={4},
  number={4},
  pages={045031},
  year={2023},
  publisher={IOP Publishing}
}

@article{incudini2024toward,
author = {Incudini, Massimiliano and Martini, Francesco and Pierro, Alessandra Di},
title = {Toward Useful Quantum Kernels},
journal = {Advanced Quantum Technologies},
pages = {1-17},
year={2024},
doi = {https://doi.org/10.1002/qute.202300298},
url = {https://onlinelibrary.wiley.com/doi/abs/10.1002/qute.202300298},
eprint = {https://onlinelibrary.wiley.com/doi/pdf/10.1002/qute.202300298},
}

@article{ragone_representation_2023,
    title = {Representation {Theory} for {Geometric} {Quantum} {Machine} {Learning}},
    author = {Ragone, Michael and Braccia, Paolo and Nguyen, Quynh T. and Schatzki, Louis and Coles, Patrick J. and Sauvage, Frederic and Larocca, Martin and Cerezo, M.},
    journal={arXiv preprint arXiv:2210.07980},
    year = {2023},
    month = {2},
    url = {http://arxiv.org/abs/2210.07980}
}

@article{bronstein_geometric_2021,
  title={Geometric deep learning: Grids, groups, graphs, geodesics, and gauges},
  author={Bronstein, Michael M and Bruna, Joan and Cohen, Taco and Veli{\v{c}}kovi{\'c}, Petar},
  journal={arXiv preprint arXiv:2104.13478},
  year={2021}
}

\end{refsection}

\end{document}